\documentclass[onecolumn]{emulateapj}
\usepackage{apjfonts}
\slugcomment{ApJ in press}
\shorttitle{Population III GRB Afterglows}
\shortauthors{Toma et al.}
\begin{document}
\title{
Population III GRB Afterglows: \\
Constraints on Stellar Masses and External Medium Densities
}
\author{Kenji Toma\altaffilmark{1,2}, 
Takanori Sakamoto\altaffilmark{3,4,5}, 
Peter M\'{e}sz\'{a}ros\altaffilmark{1,2,6}}

\altaffiltext{1}{Department of Astronomy and Astrophysics, Pennsylvania State 
University, 525 Davey Lab, University Park, PA 16802, USA (toma@astro.psu.edu; nnp@astro.psu.edu)}
\altaffiltext{2}{Center for Particle Astrophysics, Pennsylvania State University}
\altaffiltext{3}{Center for Research and Exploration in Space Science and Technology (CRESST),
NASA Goddard Space Flight Center, Greenbelt, MD 20771 (takanori@milkyway.gsfc.nasa.gov)}
\altaffiltext{4}{Joint Center for Astrophysics, University of Maryland, Baltimore Country,
1000 Hilltop Circle, Baltimore, MD 21250}
\altaffiltext{5}{NASA Goddard Space Flight Center, Greenbelt, MD 20771}
\altaffiltext{6}{Department of Physics, Pennsylvania State University}
%
%
%
\begin{abstract}
Population III stars are theoretically expected to be prominent around redshifts $z \sim 20$,
consisting of mainly very massive stars with $M_* \gtrsim 100\;M_\odot$, but there is no direct
observational evidence for these objects. They may produce collapsar gamma-ray bursts (GRBs),
with jets driven by magnetohydrodynamic processes, whose total isotropic-equivalent energy 
could be as high as $E_{\rm iso} \gtrsim 10^{57}\;$erg over a cosmological-rest-frame duration 
of $t_d \gtrsim 10^4\;$s, depending on the progenitor mass.  Here we calculate the afterglow 
spectra of such Pop.~III GRBs based on the standard external shock model, and show that they 
will be detectable with the {\it Swift} BAT/XRT and {\it Fermi} LAT instruments. 
We find that in some cases a spectral break due to electron-positron pair creation will 
be observable in the LAT energy range, which can put constraints on the ambient density 
of the pre-collapse Pop.~III star. Thus, high redshift GRB afterglow observations could be 
unique and powerful probes of the properties of Pop.~III stars and their environments. 
We examine the trigger threshold of the BAT instrument in detail, focusing on the image 
trigger system, and show that the prompt emission of Pop.~III GRBs could also be detected by BAT. 
Finally we briefly show that the late-time radio afterglows of Pop.~III GRBs for typical 
parameters, despite the large distances,  can be very bright: $\simeq 140\;$mJy at $1\;$GHz,
which may lead to a constraint on the Pop.~III GRB rate from the current radio survey 
data, and $\simeq 2.4\;$mJy at $70\;$MHz, which implies that Pop.~III GRB radio afterglows could 
be interesting background source candidates for 21~cm absorption line detections.
\end{abstract}


\section{Introduction}
\label{sec:intro}

Recent studies on cosmology and primordial star formation predict that the first 
generation of stars (population III stars) may be most prominent around $z \sim 20$,
consisting of metal-poor, mainly very massive stars (VMSs) with $M_* \gtrsim 
100\;M_\odot$ \citep[e.g.,][]{abel02,omukai03,yoshida06,ciardi05}.  These first stars 
are thought to play a significant role in setting off cosmic reionization, in the
initial enrichment of the intergalactic medium (IGM) with heavy elements, and in
seeding the intermediate and supermassive black holes (BHs) encountered in galaxies. 
The details of how these processes unfold remain elusive, since observational 
data for redshifts $z \gtrsim 6$ are very limited.

Observations of gamma-ray bursts (GRBs), however, may provide unique probes of the 
physical conditions of the universe at such redshifts. The GRB prompt emission and 
the afterglows were expected to be observable at least out to $z \gtrsim 10$, with
their redshifts being determined through the detection of a Ly$\alpha$ drop-off in 
the infrared (IR), or through redshifted atomic lines back-lighted by the afterglows
\citep[e.g.,][]{lamb00,ciardi00,gou04}. This can serve as a tracer of the history of
the cosmic star formation rate \citep[e.g.,][]{totani97,porciani01,bromm06,kistler09},
providing  invaluable information about the physical conditions in the IGM of the very 
high redshift universe \citep[e.g.,][]{barkana04,ioka05,inoue06}. Currently the most 
distant object that has been spectroscopically confirmed is GRB 090423 at $z \simeq 8.2$ 
\citep{tanvir09,salvaterra09}, and the detailed spectroscopic observation of GRB 050904 
at $z \simeq 6.3$ has put a unique upper bound on the neutral hydrogen fraction in the 
IGM at that redshift \citep{totani06,kawai06}, indicating that GRB observations are very 
promising for exploring the high-redshift universe 
\citep[see also][for GRB 080913 with $z \simeq 6.7$]{greiner09}.

In those previous papers, the GRBs arising from Population III VMSs were considered 
to have similar properties as the GRBs arising from Population I/II stars, e.g., 
they were usually assumed to have similar luminosity functions, even if perhaps extending 
to somewhat higher masses, and most importantly, their radiation properties, durations 
and spectra were modeled as being essentially similar to their lower redshift counterparts. 
However, this simplifying assumption may not be valid, as pointed out by \citet{fryer01} 
and \citet{komissarov10} (hereafter KB10). One difference is that the accretion disks 
around the much larger black holes resulting from core collapse of VMS progenitor would 
be too cool to lead to neutrino-cooled thin disks and conversion of neutrinos into 
electron-positron pairs \citep{eicher89,woosley93}. Thus, the Pop.~III GRBs are much 
likelier to be driven by MHD processes, converting the rotational energy of the central 
BH into a Poynting-flux-dominated jet \citep{blandford77}, rather than the usually assumed 
thermal-energy-dominated jets.  The total energy of a Pop.~III GRB is then proportional to 
the total disk (torus) mass, which in turn can be assumed to be proportional to the 
progenitor stellar mass, which can be much higher than that of a Pop.~I/II GRB. In addition, 
the fall-back time and/or the disk accretion time, i.e., the active duration time of a 
Pop.~III GRB jet can be much longer than that of a Pop.~I/II GRB jet, due to the larger
progenitor star.

Building on this premise, \citet{meszaros10} (hereafter MR10) proposed a possible model 
of the prompt emission and afterglow of such Pop.~III GRBs, and made rough predictions 
for their observational properties in the {\it Swift} and {\it Fermi} satellite bands. 
In this paper, we calculate in significantly more detail the very early afterglow properties of 
Pop.~III GRBs, and show that the combination of {\it Swift} and {\it Fermi} observations, 
complemented by deep IR observations of the afterglow immediately following the prompt
emission, can constrain the total isotropic-equivalent energies $(E_{\rm iso})$ of the 
Pop.~III GRBs, as well as the particle densities $n$ of their circumburst medium.  
The detection of a burst with a very high total isotropic-equivalent energy 
$E_{\rm iso} \gtrsim 10^{57}\;$erg and a very long (cosmological rest frame) duration 
$t_d \gtrsim 10^4\;$s would be strong evidence for a VMS progenitor. To constrain the 
total energy and the duration, observations of the prompt emission, whose interpretation 
is more dependent on the model details, should be complemented with observations of the 
afterglow, which is much less model-dependent.

The properties of the circumburst medium, i.e., the environments of the first stars 
prior to their collapse, have so far only been inferred from model numerical simulations, 
which differ significantly among each other. For example, the typical galactic gas 
environment could evolve as $n \propto (1+z)^4$ \citep{ciardi00}, or it might be 
approximately independent of redshift, $n \sim 0.1\;{\rm cm}^{-3}$, as a result of stellar 
radiation feedback \citep{whalen04,alvarez06}.  Observations and modeling of high-redshift GRB 
afterglows could distinguish between such numerical models.  The small number of analyses
of what are currently the most distant GRBs imply that the circumburst densities of these
high-redshift GRBs could be very different from each other, e.g. $n \simeq 10^2-10^3\;
{\rm cm}^{-3}$ for GRB 050904 with $z \simeq 6.3$ \citep{gou07}, and $n \simeq 1\;{\rm cm}^{-3}$ 
for GRB 090423 with $z \simeq 8.2$ \citep{chandra10}.  A high total energy and a high 
circumburst medium density could lead, in principle, to such a high compactness parameter of
the shocked afterglow emission region that a spectral break due to $e^+e^-$ pair production
(i.e., $\gamma\gamma$ self-absorption break) may be 
observable in the Fermi LAT energy range.  This is a unique and interesting point, which
we explore here, since the $\gamma\gamma$ self-absorption is usually not significant for 
the afterglows of Pop.~I/II GRBs \citep{zhang01}. Using this, we could constrain the circumburst
density $n$ from the observation of a $\gamma\gamma$ self-absorption break, which is a new 
method to constrain the environment of the pre-explosion GRB host galaxy.

The external shock model of the GRB afterglows seems to be robust, since it can explain many of
the late-time multi-band afterglows detected so far, and a simple extension of this model 
(e.g., continuous energy injection into the external shock) may explain many of the early-time
(observer's time $t_{\rm obs} \lesssim 1\;$hr) X-ray and optical afterglows detected by {\it Swift} 
\citep{liang07}.
More importantly, some of the very early high-energy afterglows (immediately following 
the prompt emission) recently observed by {\it Fermi} LAT are shown to be explained by this model 
\citep[e.g.,][]{kumar09,depasquale10,corsi10}.

The basic parameters of the Poynting-dominated Pop.~III GRB model are defined in 
Section~\ref{sec:jet_model}.  The very early afterglow spectrum is calculated in 
Section~\ref{sec:afterglow} (based on the standard external shock model described in Appendix),
where we discuss how to constrain $E_{\rm iso}, t_d$ and $n$ from the observations.
In Section~\ref{sec:rate} 
we deduce the effective trigger threshold of the {\it Swift} BAT instrument,
and show that the prompt emission of Pop.~III GRBs could trigger BAT.
In Section~\ref{sec:radio} we compute the late-time radio afterglow flux for a typical set 
of parameters, and evaluate the current radio survey data constraints on the Pop.~III GRB 
rate, as well as the prospects for 21~cm absorption line detection in the Pop. III GRB
radio afterglow spectra.  In Section~\ref{sec:discussion} we present a summary of our findings.

\section{Poynting-dominated Pop.~III GRB Model}
\label{sec:jet_model}

We consider VMSs rotating very fast, close to the break-up speed, as a representative case of 
Pop.~III GRB progenitor stars. Those in the $140\;M_\odot \lesssim M_* \lesssim 260\;M_\odot$ 
range are expected to explode as pair instability supernovae without leaving any compact 
remnant behind, while those in the $M_* \gtrsim 260\;M_\odot$ range are expected to undergo 
a core collapse leading directly to a central BH, whose mass would itself be hundreds of 
solar masses \citep{fryer01,heger03,ohkubo06}.  Accretion onto such BHs could lead to collapsar 
GRBs \citep{woosley93,macfadyen99}.  Prior to the collapse, the fast rotating VMSs may be 
chemically homogeneous and compact, without entering the red giant phase, so that the stellar 
radius is $R_* \simeq 10^{12}\;$cm for $M_* \simeq 10^3\;M_\odot$ 
\citep[KB10;][]{yoon06,woosley06}.

For such large BH masses $M_h \gtrsim 100 M_\odot$, the density and temperature of the 
accretion disk are too low for neutrino cooling to be important, and the low neutrino 
release from the accretion disk is insufficient to power a strong jet \citep{fryer01}.
The rate of energy deposition through this mechanism may be estimated by using the 
formula recently deduced by \citet{zalamea10} 
\begin{equation}
L_{\nu\bar{\nu}} \simeq 5 \times 10^{46}\;\dot{M}_{-1}^{9/4} M_{h,2.5}^{-3/2} \;{\rm erg}\;{\rm s}^{-1},
\end{equation}
where $\dot{M} = 0.1 \dot{M}_{-1} M_\odot\;{\rm s}^{-1}$ is the accretion rate and
$M_{h,2.5} = M_h/(10^{2.5} M_{\odot})$. This is clearly insufficient for detection from such 
high redshifts. However, strong magnetic field build-up in the accretion torus or disk  
could lead to much stronger jets, dominated by Poynting flux. Such jets will be highly 
relativistic, driven by the magnetic extraction of the rotational energy of the central BH 
through the Blandford-Znajek (BZ) mechanism \citep{blandford77}. The luminosity extracted 
from a Kerr BH with dimensionless spin parameter $a_h$ threaded by a magnetic field of 
strength $B_h$ is \citep{thorne86}
\begin{equation}
L_{\rm BZ} \approx \frac{a_h^2}{128} B_h^2 R_h^2 c,
\end{equation} 
where $R_h \approx GM_h/c^2$ is the event horizon radius of the BH. 
The dynamics of the radiatively inefficient accretion disk may be described (KB10)
through advection-dominated (ADAF) model \citep{narayan94}. For a VMS rotating at, say, half 
the break-up speed, the disk outer radius will be $R_d \simeq R_*/4$, and for a disk 
viscosity parameter $\alpha = 10^{-1} \alpha_{-1}$, the accretion time is 
\begin{equation}
t_d \simeq \frac{7}{3\alpha} \left({\frac{R_d^3}{GM_h}}\right)^{1/2} \simeq 
1.4 \times 10^4\;\alpha_{-1}^{-1} R_{*,12}^{3/2} M_{h,2.5}^{-1/2} \;{\rm s},
\label{eq:t_d}
\end{equation}
where we have defined $R_{*,12} = R_*/10^{12}\;{\rm cm}$. This gives an estimate for
both the disk lifetime and the duration of the jet, in the source frame. Given the jet 
propagation speed inside the star, $\sim 0.2 c$, deduced from magnetohydrodynamic simulations 
\citep{barkov08}, the intrinsic jet duration $t_d \simeq 10^4\;$s is sufficient to break 
through the star.  
The poloidal magnetic field strength in the disk should scale with the 
disk gas pressure, $P$, so that $B_h^2 = 8\pi P /\beta$, where $\beta = 10\;\beta_1$ is the 
magnetization parameter \citep[e.g.,][]{reynolds06}. Then we have
\begin{equation}
B_h \simeq \left(\frac{4\sqrt{14}}{3 \alpha \beta} \frac{\dot{M} c}{R_h^2}\right)^{1/2}
\simeq 6.6 \times 10^{13}\; \frac{1}{\alpha_{-1} \beta_1} M_{h,2.5}^{-1} M_{d,2.5}^{1/2} t_{d,4}^{-1/2}\; 
{\rm G}.
\label{eq:B_BZ}
\end{equation}
Combining these equations result in a jet luminosity
\begin{equation}
L_{\rm BZ} \simeq \frac{\sqrt{14}}{96} \frac{a_h^2}{\alpha \beta} \dot{M} c^2
\simeq 2.2\times10^{51}\; \frac{a_h^2}{\alpha_{-1} \beta_1} M_{d,2.5} t_{d,4}^{-1}
\;{\rm erg}\;{\rm s}^{-1},
\label{eq:L_BZ}
\end{equation}
where for the second equalities in Eqs.~(\ref{eq:B_BZ}) and (\ref{eq:L_BZ}) we have assumed 
a constant accretion rate $\dot{M} \simeq M_d/t_d$, and $M_d = 10^{2.5} M_{d,2.5} M_{\odot}$ 
is the total disk mass.

Let us assume that the factor $a_h^2/(\alpha \beta)$ is roughly constant, so that 
$L_{\rm BZ} \propto \dot{M}$. The total extracted energy during the accretion time $t_d$ is 
then $E_{\rm BZ} \simeq (\sqrt{14}/96) (a_h^2/\alpha\beta) M_d c^2$.
Assuming that the jet has an opening angle of $\theta_j = 0.1\;\theta_{j,-1}$,
we can then write the total isotropic-equivalent energy of the jet as
\begin{equation}
E_{\rm iso} \simeq 4.4 \times 10^{57}\; \frac{(1-\epsilon_{\gamma}) a_h^2}{\alpha_{-1} \beta_1} M_{d,2.5} 
\theta_{j,-1}^{-2}\;{\rm erg},
\label{eq:E_iso}
\end{equation}
where $\epsilon_\gamma$ is the radiation efficiency of the prompt emission, and 
$1-\epsilon_\gamma$ is of order of unity.  Equation~(\ref{eq:E_iso}) is also applicable to 
cases where $\dot{M}$ is not constant.  As far as 
$L_{\rm BZ} \propto t^{-q}$ (i.e., $\dot{M} \propto t^{-q}$) with $q \leq 1$, the jet 
can break out from the star and subsequently keep injecting energy into the external medium 
for the duration of the order of $t_d$.
The forward shock produced in the external medium enters a self-similar expansion phase with
total shock energy $\simeq E_{\rm iso}$ soon after $t = t_d$ \citep{blandford76}.
Interestingly, the value of $E_{\rm iso}$ for a disk mass $M_d \sim 3\;M_\odot$ is 
consistent with the observed largest value of the isotropic-equivalent $\gamma$-ray 
energy $E_{\gamma,{\rm iso}} \simeq 10^{55}\;$erg for GRB 080916C at the redshift $z \simeq 4.4$
\citep{abdo09}. 
(On the other hand, the isotropic luminosity $L_{\rm iso} \simeq 4.4 \times 10^{53}\;
(a_h^2/\alpha \beta) M_{d,2.5} t_{d,4}^{-1} \theta_{j,-1}^{-2}\;{\rm erg}\;{\rm s}^{-1}$ 
is comparable to the observed largest value.)
Thus, if we were to observe a burst at redshift $z \gtrsim 10$ with $E_{\rm iso} \gtrsim 10^{57}\;$erg, 
and with a self-similar phase starting at $t_{d,{\rm obs}} = t_d (1+z) \gtrsim 1\;$day, 
this would very likely be a burst from a Pop.~III VMS with $M_* \gtrsim 300\;M_\odot$.

\begin{figure}
\epsscale{0.75}
\plotone{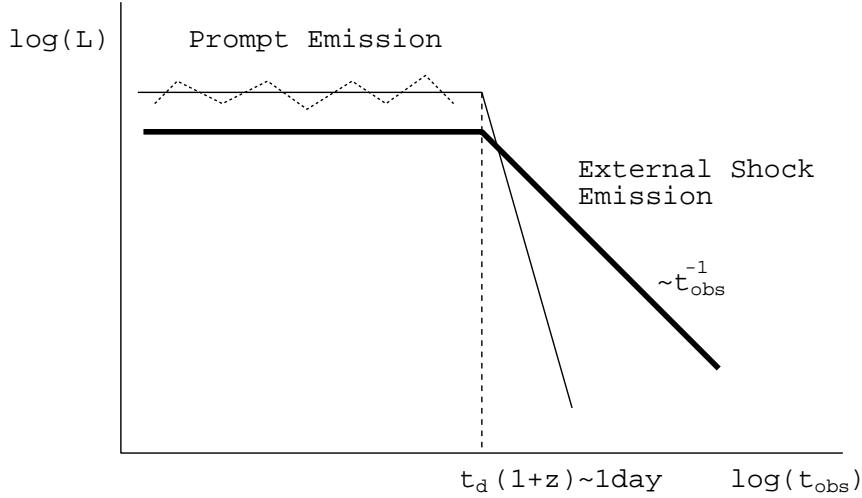}
\caption{
Schematic figure of the bolometric luminosity evolutions of the external shock emission (thick line)
and the prompt emission (thin line) as functions of time from the onset of the prompt emission.
The prompt emission may involve some variability, which is shown by the dotted line.
In this figure we assume that the jet luminosity $L_{\rm BZ}$ is constant for 
$t_{\rm obs} < t_{d,{\rm obs}} = t_d (1+z)$ 
and subsequently declines rapidly.  The external shock is in the thick-shell regime. Its luminosity is
also constant for $t_{\rm obs} < t_{d,{\rm obs}}$ and subsequently enters the self-similar phase 
$L \propto t_{\rm obs}^{-1}$.
The argument is similar for the case of the jet luminosity $L_{\rm BZ} \propto t^{-q}$ with 
$q \leq 1$, where luminosities of the prompt emission and external shock emission 
both evolve as $L \propto t_{\rm obs}^{-q}$ before $t_{d,{\rm obs}}$ and in the same way as this figure 
after $t_{d,{\rm obs}}$.
}
\label{fig:lightcurve}
\end{figure}

\section{Very Early Afterglow Spectrum}
\label{sec:afterglow}

Rough predictions for the observational properties of the afterglows of Pop.~III GRBs
were made in MR10. The external shock driven by the jet in the circumburst medium 
powers the afterglow, which can be studied independently of the prompt emission 
\citep{mesAG97,sari98,sari01}. This is true whether the jet is baryonic or Poynting-dominated, 
the jet acting simply as a piston.\footnote{The same is not true for a reverse shock, whose 
existence and properties are more dependent on the nature of the ejecta jet 
\citep{mimica09,mizuno09,lyutikov10}.  However, a reverse shock emission is most prominent 
in the low frequencies, e.g., the IR bands, while we are interested in the X-ray and 
$\gamma$-ray bands at $t_d$, so this is not considered here.}
The external shock amplifies the magnetic field in the shocked region via plasma and/or 
magnetohydrodynamic instabilities, and accelerates the electrons in the shocked region to 
a power-law energy distribution. The accelerated electrons produce synchrotron and 
synchrotron-self-Compton (SSC) radiation as an afterglow. Here we go beyond the previous
schematic outlines, and calculate the spectrum of this emission in detail, including the
Klein-Nishina as well as $e^+e^-$ pair formation effects.  

The bolometric luminosity of the external shock emission (with prompt emission light curve) 
is illustrated in Figure~\ref{fig:lightcurve}.  We focus on the external shock emission at 
the observer's time $t_{\rm obs} \simeq t_{d,{\rm obs}}$, near the beginning of the self-similar
expansion phase of the shock, when the emission is bright and may not be hidden by the prompt emission.

Calculations of the external shock emission spectrum involve the parameters 
$E_{\rm iso}$ and $t_d$, as well as the external medium number density $n$, 
the fractions $\epsilon_B$ and $\epsilon_e$ of the thermal energy in the shocked 
region that are carried by the magnetic field and the electrons, respectively, 
and the index $p$ of the energy spectrum of the accelerated electrons.
We calculate the external shock emission spectrum of a Pop.~III GRB 
at $t_{d,{\rm obs}}$ based on the standard model described in Appendix.

As introduced in Section~\ref{sec:intro}, the circumburst medium density in the very high-redshift 
universe is likely to be $n \gtrsim 0.1\;{\rm cm}^{-3}$. The microphysical parameters may be 
independent of $E_{\rm iso}, t_d,$ or $n$ as long as the shock velocity is highly relativistic,
so that $\epsilon_B, \epsilon_e$, and $p$ are thought to be similar to those for the bursts 
observed so far. Those have been constrained by fitting the late-time afterglows through models (which 
are similar to our model shown in Appendix). The parameters related to the electrons are 
constrained relatively tightly as $\epsilon_e \sim 0.1$ and $p \sim 2.3$, while those for 
the magnetic field are not so tightly constrained, although typically for many afterglows
$10^{-3} \lesssim \epsilon_B \lesssim 10^{-1}$ \citep[e.g.,][]{panaitescu02,wijers99}.

The external shock emission at $t_{d,{\rm obs}}$ will have two intrinsically different cases, 
depending on the significance of the electron-positron pair creation within the emitting region. 
These cases are characterized by a negligible pair production regime and a significant pair 
production regime, which we show examples of spectra separately below.

\subsection{Case of Negligible Pair Production}
\label{subsec:afterglow_np}

\begin{figure}
\epsscale{0.75}
\plotone{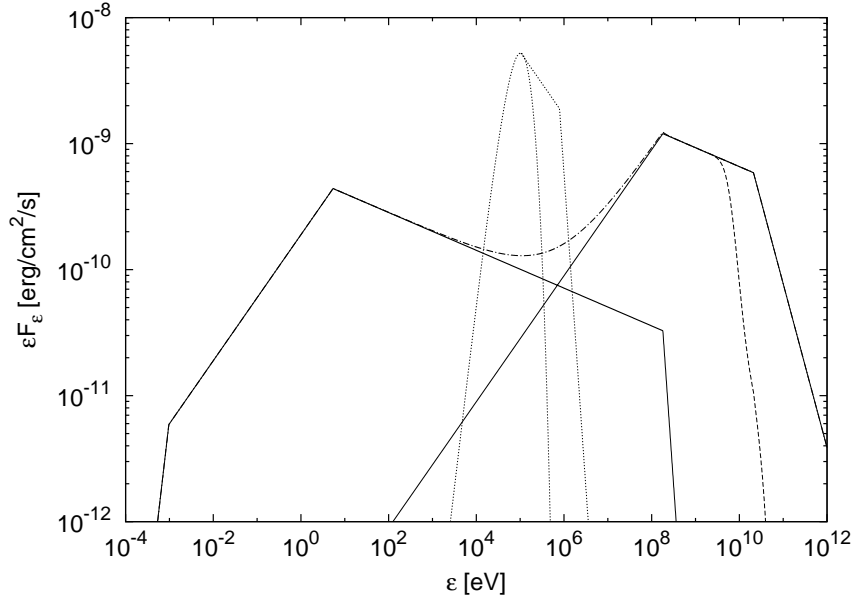}
\caption{
Example of the observer-frame spectrum of a Pop.~III GRB at the time $t_{d,{\rm obs}}$ when
the jet activity ends, for the case of negligible intra-source pair production.  The 
parameters are $E_{57.6} = t_{d,4} = n_0 = \epsilon_{B,-2} = \epsilon_{e,-1} = f(p) = 1$,
and the source redshift $1+z = 20$.
The dot-dashed line shows the external shock (afterglow) spectrum, which consists of the 
synchrotron and SSC components (solid lines). The synchrotron component 
peaks at $\varepsilon_m \simeq 5.4\;$eV, and the SSC component peaks at
$\varepsilon_m^{\rm SC} \simeq 1.8 \times 10^2\;$MeV.  The $\gamma\gamma$ self-absorption break energy 
is $\varepsilon_{\gamma\gamma} \simeq 21\;$GeV, which is larger than the SSC peak 
$\varepsilon_m^{\rm SC}$, so that most of the SSC emission escapes without being absorbed 
within the emitting region.  
The $\gamma\gamma$ absorption due to the EBL is expected to 
become significant at $\varepsilon > \varepsilon_{\rm EBL} \simeq 7\;$GeV \citep{inoue10}, 
as shown by the dashed line.  
The dotted line represents the prompt emission's dominant 
photospheric black-body component with a possible power-law extension, assuming $1+\sigma = 10$ and 
$L_{53.6} = r_{l,8} = \Gamma_l = 1$ (see Section~\ref{subsec:prompt}).
}
\label{fig:example_np}
\end{figure}

An example of the negligible pair production case is obtained for the parameters 
\begin{equation}
E_{57.6} = t_{d,4} = n_0 = \epsilon_{B,-2} = \epsilon_{e,-1} = f(p) = 1,
\end{equation}
where the notation $Q = 10^x Q_x$ in cgs units has been adopted 
($E_{57.6} = E_{\rm iso}/10^{57.6}\;{\rm erg}$).
The overall observer-frame spectrum for this case is shown in Figure~\ref{fig:example_np}
(see Appendix~\ref{subsec:case_np}).  
The synchrotron emission spectrum peaks at $\varepsilon_m \simeq 5.4\;$eV with the flux
$\varepsilon_m F_{\varepsilon_m} \simeq 4.4 \times 10^{-10}\;{\rm erg}\;{\rm cm}^{-2}\;{\rm s}^{-1}$,
having spectral breaks at synchrotron self-absorption energy $\varepsilon_a \simeq 9.7 \times 10^{-4}\;$eV 
and at energy corresponding to the maximum electron energy $\varepsilon_M \simeq 1.8 \times 10^2\;$MeV.
The SSC emission spectrum peaks at $\varepsilon_m^{\rm SC} \simeq 1.8 \times 10^2\;$MeV with the flux
$\varepsilon_m^{\rm SC} F_{\varepsilon_m}^{\rm SC} \simeq 1.2 \times 10^{-9}\;{\rm erg}\;{\rm cm}^{-2}\;
{\rm s}^{-1}$ and has a spectral break at $\gamma\gamma$ self-absorption energy 
$\varepsilon_{\gamma\gamma} \simeq 21\;$GeV. Since $\varepsilon_{\gamma\gamma} > \varepsilon_m^{\rm SC}$, 
most of the SSC emission is observed without being converted into $e^+e^-$ within the emitting region.

Even photons escaping without attenuation within the emitting region can be 
absorbed by interacting with the extragalactic background light (EBL) ($\gamma\gamma \to e^+e^-$).
\citet{inoue10} use a semi-analytic model of the evolving EBL and expect that high-energy photon 
absorption by the EBL for an arbitrary source at $z \simeq 20$ is significant at 
$\varepsilon > \varepsilon_{\rm EBL} \simeq 7\;$GeV. 
For the parameters adopted here, $\varepsilon_{\gamma\gamma}$ is larger than 
$\varepsilon_{\rm EBL}$, which precludes obtaining intrinsic information about the 
emitting region from the observation of the $\gamma\gamma$ break.

The temporal evolution of the characteristic quantities during the self-similar phase, 
i.e., at $t_{\rm obs} > t_{d,{\rm obs}}$ is obtained replacing $t_d$ by 
the variable $t_{\rm obs}$ in the model equations in Appendix, and taking all other parameters as constant.
On the other hand, for $t_{\rm obs} < t_{d,{\rm obs}}$, if the
jet luminosity evolves as $L_{\rm BZ} \propto t^{-q}$ with $q \leq 1$, one can obtain 
the temporal evolution of the characteristic quantities by taking $E_{\rm iso} 
\propto t_{\rm obs}^{1-q}$, replacing $t_d$ by $t_{\rm obs}$, and taking all the other parameters as constant. 
In the high-energy range, $\varepsilon_m^{\rm SC} < \varepsilon < \varepsilon_{\gamma\gamma}$,
as an example, we obtain $F^{\rm SC}_{\varepsilon_m^{\rm SC} < \varepsilon < 
\varepsilon_{\gamma\gamma}} \propto t_{\rm obs}^{(3/4)(2-p) - (3/8)q(p+2/3)}$ for 
$t_{\rm obs} < t_{d,{\rm obs}}$, and $\propto t_{\rm obs}^{(10-9p)/8}$ for 
$t_{\rm obs} > t_{d,{\rm obs}}$, which implies a steepening break at $t_{d,{\rm obs}}$
for $p \sim 2$ and $q < 1$. Thus one can identify the jet duration $t_{d,{\rm obs}}$ 
as the observed break time. (One can also estimate $t_{d,{\rm obs}}$ by the duration of the 
prompt emission.)

\subsection{Case of Significant Pair Production}
\label{subsec:afterglow_sp}

\begin{figure}
\epsscale{0.75}
\plotone{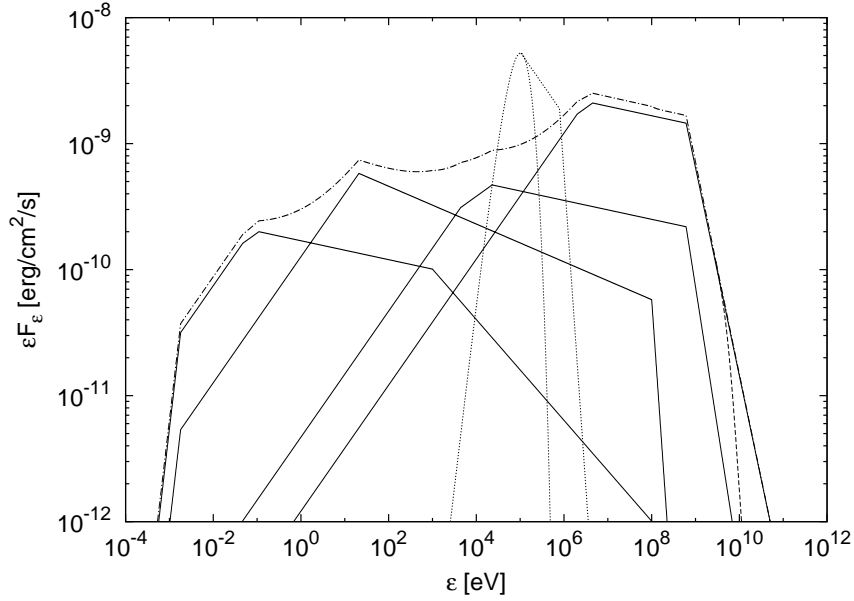}
\caption{
Example of a Pop. III GRB spectrum at time $t_d$ with significant pair production in the external shock.
The parameters are $E_{57.6} = t_{d,4} = \epsilon_{B,-2} = f(p) = 1$, $n=10^2$, and $\epsilon_{e,-1} = 2$,
and the source redshift $1+z=20$.  
The dot-dashed line shows the external shock emission spectrum, which consists of the four 
components (solid lines).  The synchrotron 
component of the original electrons peaks at $\varepsilon_m \simeq 21\;$eV.  The peak of the 
SSC emission of the original electrons (not shown in this figure) $\varepsilon_m^{\rm SC} 
\simeq 9.2\times10^2\;$MeV is above the $\gamma\gamma$ self-absorption break energy 
$\varepsilon_{\gamma\gamma} \simeq 6.1\times10^2\;$MeV, so that most of the SSC emission is absorbed
within the emission region.
The pairs emit synchrotron and SSC radiations peaking at $\varepsilon_{\pm,p} \simeq 
0.11\;$eV and $\varepsilon_{\pm,p}^{\rm SC} \simeq 23\;$keV, respectively.
The IC-scattered original synchrotron emission by the pairs and the IC-scattered pair 
synchrotron emission by the original electrons have spectra with similar characteristic 
energies (including the peak energies at $\varepsilon_m^{\rm IC} = 
\varepsilon_{\pm,p}^{\rm IC} \simeq 4.5\;$MeV) and different flux normalizations, which 
have been superposed in this figure.  
The $\gamma\gamma$ absorption due to the EBL is expected to be significant above 
$\simeq 7\;$GeV \citep{inoue10}, as shown by the dashed line.  The dotted line represents the 
prompt emission's dominant photospheric black-body component with a possible power-law extension, 
assuming $1+\sigma = 10$ and $L_{53.6} = r_{l,8} = \Gamma_l = 1$ (see Section~\ref{subsec:prompt}).
}
\label{fig:example_sp}
\end{figure}

An example of the case of significant pair production is obtained with the parameter set
\begin{equation}
E_{57.6} = t_{d,4} = \epsilon_{B,-2} = f(p) = 1, ~~~ n_0 = 10^2, ~~~ \epsilon_{e,-1} = 2.
\end{equation}
The overall observer-frame spectrum for this case is shown in Figure~\ref{fig:example_sp}
(see Appendix~\ref{subsec:case_sp}).  
The synchrotron emission spectrum of the original electrons peaks at $\varepsilon_m \simeq 21\;$eV 
with the flux $\varepsilon_m F_{\varepsilon_m} \simeq 
5.8 \times 10^{-10}\;{\rm erg}\;{\rm cm}^{-2}\;{\rm s}^{-1}$, having the maximum energy
$\varepsilon_M \simeq 1.0 \times 10^2\;$MeV.  The $\gamma\gamma$ self-absorption energy is 
$\varepsilon_{\gamma\gamma} \simeq 6.1 \times 10^2\;$MeV, and thus most of the SSC emission 
with the spectral peak $\varepsilon_m^{\rm SC} \simeq 9.2 \times 10^2\;$MeV with the peak flux 
$\varepsilon_m^{\rm SC} F_{\varepsilon_m}^{\rm SC}
\simeq 2.7 \times 10^{-9}\;{\rm erg}\;{\rm cm}^{-2}\;{\rm s}^{-1}$
is absorbed within the emitting region.  The created pairs emit synchrotron emission peaking 
at $\varepsilon_{\pm,p} \simeq 0.11\;$eV and SSC emission
peaking at $\varepsilon_{\pm,p}^{\rm SC} \simeq 23\;$keV. In addition to these,
the pairs Inverse Compton (IC)-scatter the original electron synchrotron emission and the original 
electrons IC-scatter the pair synchrotron emission, which have similar spectra peaking at 
$\varepsilon_{\pm,p}^{\rm IC} = \varepsilon_m^{\rm IC} \simeq 4.5\;$MeV
with different flux values, and they have been superposed in Figure~\ref{fig:example_sp}.

In this case, we can measure $\varepsilon_{\gamma\gamma}$, since this is well below the expected
EBL cut-off, so one would be able to draw inferences about the source parameters from the 
$\gamma\gamma$ self-absorption break (see next section for details). 

The temporal evolution of the flux in the LAT energy range will be
$F_{\varepsilon_{\pm,p}^{\rm IC} < \varepsilon < \varepsilon_{\gamma\gamma}} 
\propto t_{\rm obs}^{-(7/8)(p-2)-(1/16)q(7p+2)}$ for $t_{\rm obs} < t_{d,{\rm obs}}$ and
$\propto t_{\rm obs}^{(26-21p)/16}$ for $t_{\rm obs} > t_{d,{\rm obs}}$, which implies 
a steepening break at $t_{d,{\rm obs}}$.
The lightcurve well after $t_{d,{\rm obs}}$, however, may be complicated. The SSC component could become
dominant at later times, since the SSC energy evolves as $\varepsilon_m^{\rm SC} \propto 
t_{\rm obs}^{-9/4}$ and the $\gamma\gamma$ self-absorption energy evolves as $\varepsilon_{\gamma\gamma} 
\propto t_{\rm obs}^{(3/4)-(2/p)}$ for $t_{\rm obs} > t_{d,{\rm obs}}$.

\subsection{Constraints on $E_{\rm iso}$ and $n$ from Observations}
\label{subsec:constraints}

We have shown two typical cases of the external shock emission of Pop.~III GRBs, one
being the case of negligible pair production, and the other being the case of significant pair 
production for which the cascade process stops when the first generation pairs are created.
There may be cases where the cascade process can create second (or higher) generation pairs,
e.g., for larger $\epsilon_e$ and/or larger external $n$ (in the above example we used
$\epsilon_e = 0.2$ and $n=10^2\;{\rm cm}^{-3}$). In any case, the important point is that one 
will be able to detect a spectral break at energy $\varepsilon_{\gamma\gamma}$ due to pair 
creation {\it within the emission region} in the 
{\it Fermi} LAT energy range, 50 MeV - 30 GeV. This is a unique feature of GRB afterglows 
with very large $E_{\rm iso}$ (as expected for Pop. III GRB) and modest to large external 
density $n$.  Equation~(\ref{eq:gammagamma}) or (\ref{eq:compactness}) indicate that 
larger $E_{\rm iso}$ and $n$ lead to smaller $\varepsilon_{\gamma\gamma}$, increasing its
diagnostic value. This is in contrast to the usual case of Pop.~I/II GRBs, where the 
$\gamma\gamma$ self-absorption energy is not relevant for observations \citep{zhang01}.

\begin{figure}
\epsscale{0.75}
\plotone{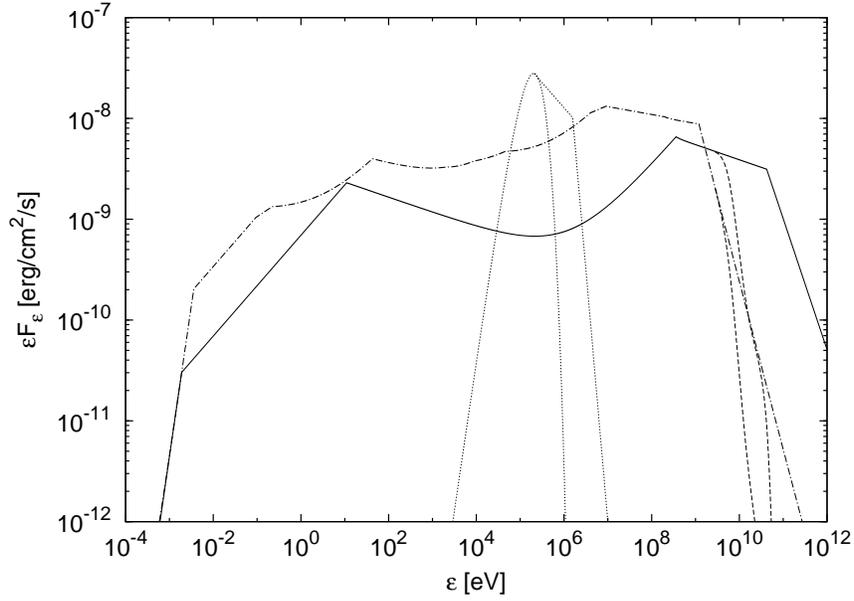}
\caption{
The spectra for a lower redshift $1+z=10$, showing the total external shock emission in the 
case of negligible pairs (solid curve; c.f. Figure~\ref{fig:example_np}) and in the case
of significant pairs (dot-dashed curve; c.f. Figure~\ref{fig:example_sp}), with the 
prompt photospheric emission (dotted curve). The EBL $\gamma\gamma$ cutoff is shown by
the dashed lines, with a cutoff energy $\varepsilon_{\rm EBL}$ assumed to be similar to
that in the $1+z=20$ figures.
}
\label{fig:example_z10}
\end{figure}

We can estimate the detection thresholds in the high energy ranges from the joint observation 
of GRB 090510 by {\it Swift} and {\it Fermi} \citep{depasquale10}.
This indicates that the thresholds of the 1-day averaged $\varepsilon F_{\varepsilon}$ flux
are $\sim 6 \times 10^{-15}\;{\rm erg}\;{\rm cm}^{-2}\;{\rm s}^{-1}$ in the XRT energy range
0.3 - 10 keV, $\sim 3 \times 10^{-10}\;{\rm erg}\;{\rm cm}^{-2}\;{\rm s}^{-1}$ in the BAT 
energy range 15 - 150 keV, and $\sim 3 \times 10^{-11}\;{\rm erg}\;{\rm cm}^{-2}\;{\rm s}^{-1}$ 
in the LAT energy range 50 MeV - 30 GeV. 
Compared to the results shown in Figures~\ref{fig:example_np} and \ref{fig:example_sp}, it appears 
that the thresholds of the XRT and LAT are thus sufficiently low, and the BAT is marginally low
only for the case of Figure~\ref{fig:example_sp}, 
to observe the high-energy spectrum of the external shock emission of Pop. III GRB. 
Furthermore, for both cases of Figures ~\ref{fig:example_np} and \ref{fig:example_sp}, 
we find that the very high energy
emission at $\varepsilon \gtrsim 10\;$GeV could be detected with next generation facility such as 
Cherenkov Telescope Array (CTA)\footnote{http://www.cta-observatory.org.}, which will have a 
threshold of 1-day averaged $\varepsilon F_{\varepsilon}$ flux
$\sim 10^{-12}\;{\rm erg}\;{\rm cm}^{-2}\;{\rm s}^{-1}$ for $\varepsilon \gtrsim 10\;$GeV,
although this significantly depends on the EBL attenuation for individual burst.
In Figure~\ref{fig:example_z10} we show, for reference, the similar results for a lower redshift
of $1+z=10$, using the same values of the other parameters as for the previous two figures.
The above statement is also applicable to this case. The EBL attenuation effect
for $1+z=10$ is expected to be similar to that
for $1+z=20$ since the EBL intensity declines at $z \gtrsim 10$ \citep{inoue10}.

One of the main questions that will be asked, if and when the redshift of a burst is 
determined to be $z \gtrsim 10$ (e.g. by observation of the Ly$\alpha$ cutoff at 
IR frequencies), is whether this burst is produced by a Pop.~III VMS or not. 
An effective way to pinpoint a Pop.~III progenitor is examine whether the afterglow
spectrum from its surrounding medium is devoid of metals through high resolution IR 
and X-ray spectroscopy by ground based facilities and/or future space experiments.
Here we detailedly discuss another way by estimating the duration and total energy of the jet
through the X-ray and $\gamma$-ray observations of the afterglow and/or prompt emission.
The jet duration timescale $t_d$ can be estimated from the steepening of the afterglow light 
curve and/or the end of the prompt emission 
(see Sections~\ref{subsec:afterglow_np} and \ref{subsec:afterglow_sp}, and Figure~\ref{fig:lightcurve}).
A lower bound on the total isotropic-equivalent 
energy $E_{\rm iso}$ can be estimated from the observed flux level in a specific energy range.
Thus, if one obtains an $E_{\rm iso} \gtrsim 10^{57}\;$erg and $t_d \gtrsim 10^4\;$s, 
this burst is almost certainly bound to be a Pop.~III GRB. 

One could in principle attempt to constrain the electron spectral index $p$ as well, using 
the observed photon spectral index, and the four physical parameters $E_{\rm iso}$, $n$, 
$\epsilon_B$, and $\epsilon_e$, by measuring the observables $\varepsilon_m$, $F_{\varepsilon_m}$, 
$\varepsilon_m^{\rm SC}$ and $F_{\varepsilon_m}^{\rm SC}$ with the {\it Swift} XRT,  the
{\it Fermi} LAT, and a possible IR detection. However, we have shown that there are some 
cases in which the spectral peak in the LAT energy range corresponds to 
$\varepsilon_{\pm,p}^{\rm IC}$. In such cases one cannot easily distinguish between 
the regime of negligible pair production and that of significant pair production, so that 
one would not be able to uniquely constrain all four physical parameters.  Furthermore, 
the prompt emission may last until $t_{\rm obs} \simeq t_{d,{\rm obs}}$, which could hide the peak of the 
external shock synchrotron or IC/SSC emission, reducing the number of observables of the 
external shock emission. 

We consider now the case in which the external shock emission is observed in the 
LAT energy range, without being hidden by the prompt emission component, and show how
this can constrain $E_{\rm iso}$ as well as $n$.
In this case we have two observables, the flux at some energy in the 
LAT energy range, $\varepsilon_L F_{\varepsilon_L}$, and the $\gamma\gamma$
self-absorption break energy $\varepsilon_{\gamma\gamma}$, which is identifiable if it is 
well below $\varepsilon_{\rm EBL}$.

The LAT flux can put a lower limit on $E_{\rm iso}$, modulo the uncertainty on 
$\epsilon_e$.  The LAT flux should be lower than 
$\varepsilon_m^{\rm SC} F_{\varepsilon_m}^{\rm SC}$ or $\varepsilon_{\pm,p}^{\rm IC}
F_{\varepsilon_{\pm,p}}^{\rm IC}$ for the case of $\epsilon_e \gtrsim \epsilon_B$
or $\varepsilon_m F_{\varepsilon_m}$ for the case of $\epsilon_e \ll \epsilon_B$. 
In any case, we have that
$\varepsilon_L F_{\varepsilon_L} \lesssim \epsilon_e E_{\rm iso} t_d^{-1}
(p-2)/[4\pi d_L^2(p-1)]$. We may approximate $p$ to be $2 \lambda_L$, where $\lambda_L$
is the measured spectral index in the LAT energy range. This leads to
\begin{equation}
E_{\rm iso} \gtrsim 2 \times 10^{57}\; \epsilon_{e,-1}^{-1} 
\left(\frac{\varepsilon_L F_{\varepsilon_L}}{10^{-9}\;{\rm erg}\;{\rm cm}^{-2} 
{\rm s}^{-1}}\right) t_{d,4} f_L(\lambda_L) d_{L,20}^2 \;{\rm erg},
\end{equation}
where $f_L(\lambda_L) = (2/7)(\lambda_L-0.5)/(\lambda_L-1)$, and 
the luminosity distance $d_L$ is normalized by the value for $1+z=20$,
$6.7 \times 10^{29}\;$cm.
This bound can be compared with the total isotropic-equivalent energy of the prompt 
emission, $E_{\gamma,{\rm iso}}$.

From an observation of the $\gamma\gamma$ self-absorption break in the LAT band, we 
can then constrain the density of the medium around the Pop.~III 
star before its collapse.  The EBL cutoff energy $\varepsilon_{\rm EBL}$ can be 
estimated by some EBL models \citep[e.g.,][]{inoue10} when we have the source redshift $z$.
If we detect a spectral break which is well below the values of $\varepsilon_{\rm EBL}$ expected for
practically all EBL models, that is likely to be $\varepsilon_{\gamma\gamma}$.  
By using this, we can constrain the bulk Lorentz factor of the emitting region $\Gamma_d$
\citep[see Appendix and][]{zhang01,lithwick01}. For the photons at $\varepsilon_{\gamma\gamma}$ in 
the LAT energy range, the main target photons have energies at $\varepsilon_{at} \sim 
\Gamma_d^2 m_e^2 c^4/[(1+z)^2 \varepsilon_{\gamma\gamma}] \simeq 7\;(\Gamma_d/10^2)^2 
(\varepsilon_{\gamma\gamma}/1\;{\rm GeV})^{-1} [(1+z)/20]^{-2}\;$keV. Thus we can estimate the target 
photon number density by using the {\it Swift} XRT data. 
The equation for the optical depth $\tau_{\gamma\gamma}(\varepsilon_{\gamma\gamma}) \simeq (\sigma_T/10) 
(d_L^2/r_d^2) F_{\nu_X} (\varepsilon_{at}/\varepsilon_X)^{-\lambda} t_d/[h(1+z)]=1$, 
where $\varepsilon_X$, $F_{\nu_X}$ and $\lambda$ are the observed X-ray energy, 
flux and spectral index, with the equation for the emission radius $r_d \simeq c \Gamma_d^2 t_d$, leads to
\begin{eqnarray}
\Gamma_d \simeq
60\;(20)^{\frac{\lambda-1.2}{\lambda+2}}\;
\left(\frac{F_{\nu_X}}{10^{-4}\;{\rm Jy}}\right)^{\frac{1}{2\lambda +4}}
\left(\frac{\varepsilon_X}{1\;{\rm keV}}\right)^{\frac{\lambda}{2\lambda +4}}
\left(\frac{\varepsilon_{\gamma\gamma}}{1\;{\rm GeV}}\right)^{\frac{\lambda}{2\lambda +4}} 
t_{d,4}^{\frac{-1}{2\lambda +4}} d_{L,20}^{\frac{1}{\lambda +2}} 
[(1+z)/20]^{\frac{2\lambda -1}{2\lambda +4}}.
\end{eqnarray} 
By using $E_{\rm iso} \simeq 4\pi r_d^3 \Gamma_d^2 n m_p c^2$, we have an estimate of 
$E_{\rm iso}/n$.  Combining it with the above lower limit on $E_{\rm iso}$, we can put a 
lower limit on $n$,
\begin{eqnarray}
n &\gtrsim& 40\;(20)^{\frac{8(1.2-\lambda)}{\lambda+2}} \;\epsilon_{e,-1}^{-1}
\left(\frac{F_{\nu_X}}{10^{-4}\;{\rm Jy}}\right)^{\frac{-4}{\lambda +2}}
\left(\frac{\varepsilon_X}{1\;{\rm keV}}\right)^{\frac{-4\lambda}{\lambda +2}} 
\left(\frac{\varepsilon_{\gamma\gamma}}{1\;{\rm GeV}}\right)^{\frac{-4\lambda}{\lambda +2}}
\left(\frac{\varepsilon_L F_{\varepsilon_L}}
{10^{-9}\;{\rm erg}\;{\rm cm}^{-2}\;{\rm s}^{-1}}\right) \nonumber 
\\ &&\times\; t_{d,4}^{\frac{-2\lambda}{\lambda +2}} f_L(\lambda_L)
d_{L,20}^{\frac{2\lambda -4}{\lambda +2}} 
[(1+z)/20]^{\frac{4(1- 2\lambda)}{\lambda +2}} 
\;{\rm cm}^{-3}.
\end{eqnarray}
Even if the prompt emission hides the external shock X-ray emission, taking 
the prompt emission X-ray flux as $F_{\nu_X}$ may provide us with a 
good estimate for a lower limit on $n$.

In principle, one could have a situation where  $\epsilon_e \ll \epsilon_B$, although this
appears to be rare. In this case the SSC component would be dim, and any high-energy cutoff 
(or break) is due to a synchrotron maximum energy $\varepsilon_M$.  Since 
$\varepsilon_{\gamma\gamma} \propto t_{\rm obs}^{-0.1}$ for $p \sim 2$ and 
$\varepsilon_M \propto t_{\rm obs}^{-3/8}$,
we can distinguish the cutoff origins. If the cutoff is $\varepsilon_M$, we can compute the bulk
Lorentz factor by using Equation~(\ref{eq:epsilon_M}) and a lower limit on $n$.

Above we have considered the cases in which the afterglow emission can be well observed
in the XRT and LAT energy ranges. It would be useful to examine for what ranges of parameters
the emission cannot be well observed. At $\varepsilon > \varepsilon_m$, we have a rough but simple 
estimate of the flux as $\varepsilon F_{\varepsilon} \sim \epsilon_e E_{\rm iso} t_d^{-1} 
(p-2)/[4\pi d_L^2 (p-1)] \propto \epsilon_e L_{\rm iso} d_L^{-2} f(p)$, where 
$L_{\rm iso} = E_{\rm iso} t_d^{-1} \propto M_d M_h^{1/2} R_*^{-3/2} \theta_j^{-2}$, and
the flux is found to be weakly dependent on $\epsilon_B$ or $n$.
Although we have still several free parameters, we may examine the rough parameter dependence
of the flux level. As an example of smaller progenitor mass $M_* \sim 100\;M_{\odot}$ (which 
corresponds to $M_{d,2.5} \simeq M_{h,2.5} \simeq 0.1$), if we assume $R_{*,12} \simeq 0.5$ and 
$\theta_{j,-1} \simeq \epsilon_{e,-1} \simeq f(p) \simeq d_{L,20} \simeq 1$,
we have $\varepsilon F_{\varepsilon} \sim 1 \times 10^{-10}\;{\rm erg}\;{\rm cm}^{-2}\;{\rm s}^{-1}$
at $t_{d,{\rm obs}} \sim 3\;{\rm day}$, which is still well above the XRT threshold, but below the BAT 
threshold, for the 1-day integration of the flux (shown above). This flux is marginally above the 
LAT threshold, but the spectral break at $\varepsilon_{\gamma\gamma}$ may be difficult to be 
clearly identified. However, note that the flux could be much higher than this value, depending on 
the poorly constrained parameters $R_*$, $\theta_j$, and $\epsilon_e$. 
The XRT instrument seems to be very powerful to 
observe the emission for even smaller mass progenitors, but anyway in order to identify the direction
to the GRBs on the sky, the emission flux of the afterglow or the prompt emission has to be 
sufficiently high to trigger the BAT instrument. This issue is discussed below.

\section{Detectability and Rate of Pop.~III GRBs}
\label{sec:rate}

\subsection{{\it Swift} BAT detection threshold}
\label{subsec:BAT}

Instruments with large field of view need to be triggered by a GRB in order to observe them 
from early times.  Here we investigate the detection threshold of the {\it Swift} BAT in 
detail to address whether Pop.~III GRBs can trigger the BAT, by focusing on the ``image trigger'' mode.  

The regular ``rate trigger'' mode looks for a rate increase in the light curves.  The rate trigger is 
thus sensitive to a burst which is variable on a relatively short time scale.  On the other 
hand, the image trigger searches for a burst by creating sky images every $64\;$s in the $15-50\;$keV 
band.  The image trigger is purely based on whether 
a new source is found in the sky image or not in the given interval, without looking for a rate 
increase in the light curves.  Generally it is not simple to 
find the detection threshold of the BAT instrument, because 494 different trigger criteria 
(e.g. different energy bands, time scales and combinations of detectors) have been running 
on-board.  We are able to estimate a reasonable detection threshold, however, by investigating
only the BAT image triggered GRBs, since
the image trigger with the 64 s integration is based on a criterion 
with a fixed time-scale and energy band.

\begin{figure}
\includegraphics[width=12cm,angle=-90,scale=0.7]{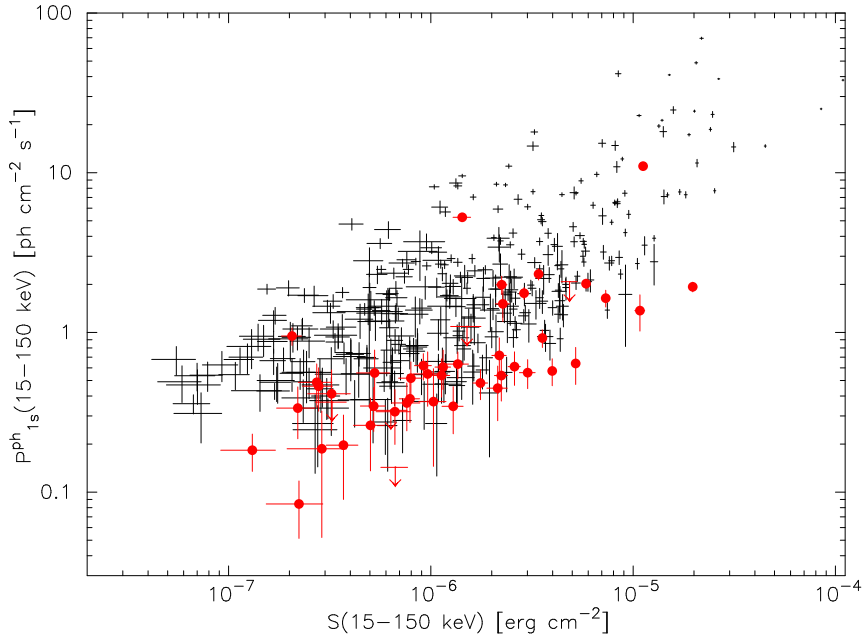}
\caption{The 1-s peak photon flux in the 15-150 keV band versus the fluence in the 15-150 keV band 
for the Swift GRBs in the BAT2 catalog \citep{sakamoto10}
overlaid with 64s image triggered GRBs (red).}
\label{fig:bat_s_p}
\end{figure}

There are some other reasons for focusing on the image triggered bursts to obtain a reasonable
detection threshold of BAT.
Figure~\ref{fig:bat_s_p} shows a diagram of the peak photon fluxes and the fluences in 
the $15-150\;$keV band of {\it Swift} GRBs in the BAT2 catalog which includes $64\;$s image 
triggered GRBs \citep{sakamoto10}.  The BAT GRBs found by the 64 s image trigger have systematically 
weaker peak photon flux (but similar fluence) compared to those GRBs found by the rate triggers.  
Furthermore, the prompt emission and external shock emission of Pop.~III GRBs are likely to 
have a lower peak flux and a longer duration, and to be less variable due to intrinsic property
and high-redshift time stretching.  The image trigger mode may be more sensitive to such emission
than the rate trigger mode. Indeed, the high-redshift GRB 050904 \citep[with $z \simeq 
6.3$,][]{kawai06} and the low-luminosity GRB 060218 \citep{campana06,toma07} were image triggered 
bursts.  Thus the estimate of the threshold of the image trigger mode may be relevant especially for 
Pop.~III GRBs. (Note, however, that GRB 090423 \citep[with $z \simeq 8.2$,][]{tanvir09,salvaterra09} 
and GRB 080913 \citep[with $z \simeq 6.7$,][]{greiner09} are the rate triggered bursts.)

\begin{figure}
\includegraphics[width=12cm,angle=-90,scale=0.7]{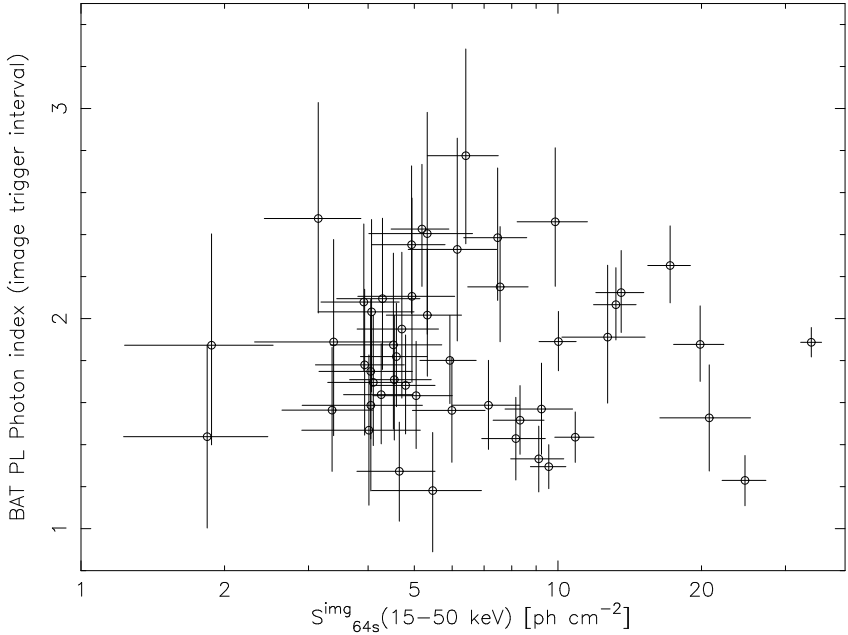}
\caption{The photon index versus the photon fluence in the 15-50 keV band in the image interval 
of the 64s image triggered GRBs.  The spectral model is a simple power-law model.}
\label{fig:bat_ph_fluence_phindex}
\end{figure}

There are 72 GRBs (out of 467 GRBs) detected by the image trigger in the 2nd BAT GRB catalog.
Out of 72 imaged triggered GRBs, there are 50 GRBs found by the image trigger with an integration 
time of $64\;$s,\footnote{Although the image trigger is basically producing images every 64 s, the image 
triggered interval could be longer than $64\;$s when the event has also been triggered by the rate 
trigger during the image triggered interval.}  
which have been shown in the red points in Figure~\ref{fig:bat_s_p}.
We focus on this sample of 50 GRBs to estimate the BAT 
detection threshold.  To understand the BAT detection threshold for the image trigger GRBs, we 
create the spectrum using the $64\;$s image trigger interval, and extract the photon fluence of 
the interval in the $15-50\;$keV band (same energy band of the image trigger) and the best fit 
photon index based on a simple power-law model.
The image trigger interval has a duration of 64 s starting from the BAT trigger time.  
Figure~\ref{fig:bat_ph_fluence_phindex} shows the distribution of the photon index and the 
photon fluence in the $15-50\;$keV band for a 64 s image interval of 50 image trigger GRBs.  
The imaged fluence in the $15-50\;$keV band does not have a strong dependence on the observed 
spectrum.  Therefore, we conclude that the photon fluence threshold of the BAT $64\;$s image 
trigger is $\sim 1\;{\rm ph}\;{\rm cm}^{-2}$ in the $15-50\;$keV band, corresponding to a photon 
fluence threshold in the $15-150\;$keV band, of $\sim 2\;{\rm ph}\;{\rm cm}^{-2}$.
This corresponds to the averaged photon flux $\sim 0.03\;{\rm ph}\;{\rm cm}^{-2}\;{\rm s}^{-1}$
in the $15-150\;$keV band.
Note that this is the minimum averaged flux in the first $64\;$s interval of the image triggered
GRBs, which is different from their minimum peak flux for the total observed duration,
shown in Figure~\ref{fig:bat_s_p}. 

\begin{table}
\caption{Prompt emission parameters for 33 image trigger GRBs with no optical counterpart. 
$T_{90}$ is $T_{90}$ duration of the burst.  S(15-150) is the fluence in the 15-150 keV band.  
P$^{1s}_{\rm ph}$(15-150) is the 1-s peak photon flux in the 15-150 keV band.  
F$^{\rm IT}$(15-50) is the energy flux measured in the image triggered 64 s interval in the 15-50 
keV band.  PL PhIndex$^{\rm IT}$ is the photon index based on a simple power-law model to the 
spectrum of the 64 s image triggered interval. \label{tab:bat_33_IT_grb}}
\begin{center}
{\scriptsize
\begin{tabular}{lccc|cc}\hline
GRB       & $T_{90}$ & S(15-150)        & P$^{1s}_{\rm ph}$(15-150)    & F$^{\rm IT}$(15-50) & PL PhIndex$^{\rm IT}$ \\
              & (s)          & (erg cm$^{-2}$) & (ph cm$^{-2}$ s$^{-1}$) & (erg cm$^{-2}$ s$^{-1}$)               &               \\\hline
050714B & 46.9      & $(5.3 \pm 1.0) \times 10^{-7}$ & $0.6 \pm 0.2$ & $(6.0 \pm 1.0) \times 10^{-9}$ & $2.5 \pm 0.4$\\
050803   & 88.1      & $(2.2 \pm 0.1) \times 10^{-6}$ & $1.0 \pm 0.1$ & $(4.2 \pm 0.8) \times 10^{-9}$ & $1.6 \pm 0.3$\\
050916   & 80.0        & $(1.1 \pm 0.2) \times 10^{-6}$ & $0.5 \pm 0.2$ & $(4.3 \pm 0.7) \times 10^{-9}$ & $1.8 \pm 0.3$\\
050922B & 156.3    & $(2.4 \pm 0.4) \times 10^{-6}$ & $1.1 \pm 0.4$ & $(7.2 \pm 1.0) \times 10^{-9}$ & $2.0 \pm 0.3$\\
051001   & 190.6    & $(1.8 \pm 0.1) \times 10^{-6}$ & $0.5 \pm 0.1$ & $(2.6 \pm 0.6) \times 10^{-9}$ & $2.0 \pm 0.3$\\
051213   & 71.1      & $(8.0 \pm 1.0) \times 10^{-7}$ &  $0.5 \pm 0.1$ & $(3.4 \pm 0.6) \times 10^{-9}$ & $1.6 \pm 0.3$\\
051221B & 39.9     & $(9.1 \pm 1.0) \times 10^{-7}$ & $0.6 \pm 0.2$ & $(2.4 \pm 0.5) \times 10^{-9}$ & $1.7 \pm 0.3$\\
060211A & 118.2   & $(1.6 \pm 0.1) \times 10^{-6}$ & $0.4 \pm 0.1$ & $(3.0 \pm 0.3) \times 10^{-9}$ & $1.4 \pm 0.2$\\
060319   & 8.9       & $(2.4 \pm 0.3) \times 10^{-7}$ & $1.1 \pm 0.1$ & $(2.3 \pm 0.4) \times 10^{-9}$ & $2.0 \pm 0.4$\\
060413   & 117.3   & $(3.6 \pm 0.1) \times 10^{-6}$ & $0.9 \pm 0.1$ & $(2.3 \pm 0.5) \times 10^{-9}$ & $1.6 \pm 0.3$\\
060427   & 62.0    & $(5.0 \pm 0.9) \times 10^{-7}$ & $0.3 \pm 0.1$ & $(3.1 \pm 0.5) \times 10^{-9}$ & $2.0 \pm 0.4$\\
060516   & 161.2  & $(1.0 \pm 0.2) \times 10^{-6}$ & $0.4 \pm 0.2$ & $(3.4 \pm 0.6) \times 10^{-9}$ & $2.0 \pm 0.3$\\
060728$^{\dagger}$   & -         & $(2.2 \pm 0.7) \times 10^{-7}$ & $0.08 \pm 0.02$ & $(1.2 \pm 0.4) \times 10^{-9}$ & $1.4 \pm 0.4$\\
060923C & 67.4 & $(1.6 \pm 0.2) \times 10^{-6}$ & $0.9 \pm 0.3$ & $(1.1 \pm 0.1) \times 10^{-8}$ & $2.2 \pm 0.2$\\
061027$^{\dagger}$   & -         & $(2.9 \pm 0.9) \times 10^{-7}$ & $0.2 \pm 0.1$ & $(2.2 \pm 0.7) \times 10^{-9}$ & $1.9 \pm 0.5$\\
061028 & 105.6 & $(9.5 \pm 1.7) \times 10^{-7}$ & $0.6 \pm 0.2$ & $(3.0 \pm 0.6) \times 10^{-9}$ & $1.9 \pm 0.4$\\
070126$^{\dagger}$ & - & $(1.3 \pm 0.4) \times 10^{-7}$ & $0.18 \pm 0.04$ & $(1.2 \pm 0.4) \times 10^{-9}$ & $1.9 \pm 0.5$\\
070429A & 168.0 & $(9.4 \pm 1.4) \times 10^{-7}$ & $0.4 \pm 0.1$ & $(3.9 \pm 0.6) \times 10^{-9}$ & $2.2 \pm 0.3$\\
070520A & 71.0  & $(5.0 \pm 1.0) \times 10^{-7}$ & $0.3 \pm 0.1$ & $(2.5 \pm 0.5) \times 10^{-9}$ & $1.9 \pm 0.3$\\
070704   & 377.6 & $(5.9 \pm 0.3) \times 10^{-6}$ & $2.0 \pm 0.1$ & $(1.1 \pm 0.1) \times 10^{-8}$ & $2.2 \pm 0.2$\\
070920A & 51.5 & $(5.2 \pm 0.7) \times 10^{-7}$ & $0.3 \pm 0.1$ & $(2.9 \pm 0.4) \times 10^{-9}$ & $1.6 \pm 0.2$\\
071018$^{\star}$   & 288.0 & $(1.1 \pm 0.2) \times 10^{-6}$ & $<0.2$ & $(1.0 \pm 0.2) \times 10^{-9}$ & $1.6 \pm 0.3$\\
071021   & 228.7 & $(1.4 \pm 0.2) \times 10^{-6}$ & $0.6 \pm 0.1$ & $(2.6 \pm 0.5) \times 10^{-9}$ & $1.8 \pm 0.4$\\
071028A$^{\star}$ & 33.0       & $(3.3 \pm 0.6) \times 10^{-7}$ & $<0.4$ & $(3.0 \pm 0.5) \times 10^{-9}$ & $1.8 \pm 0.2$\\
080207$^{\ddagger}$  & $>290$ & -                                           & -          & $(6.7 \pm 0.5) \times 10^{-9}$ & $1.3 \pm 0.1$\\
080325$^{\star}$   & 162.8 & $(4.9 \pm 0.4) \times 10^{-6}$ & $<2.0$ & $(1.4 \pm 0.3) \times 10^{-8}$ & $1.5 \pm 0.3$\\
081017$^{\star}$   & $>320$ & $(1.4 \pm 0.2) \times 10^{-6}$ & $0.2 \pm 0.1$ & $(1.9 \pm 0.3) \times 10^{-9}$ & $1.6 \pm 0.2$\\
081022   & 157.6 & $(2.6 \pm 0.2) \times 10^{-6}$ & $0.6 \pm 0.1$ & $(7.5 \pm 0.6) \times 10^{-9}$ & $1.4 \pm 0.1$\\
090308   & 25.1 & $(2.2 \pm 0.5) \times 10^{-7}$ & $0.3 \pm 0.1$ & $(3.1 \pm 0.5) \times 10^{-9}$ & $2.4 \pm 0.4$\\
090401A & 117.0 & $(1.12 \pm 0.03) \times 10^{-5}$ & $11.0 \pm 0.4$ & $(6.3 \pm 0.9) \times 10^{-9}$ & $1.6 \pm 0.2$\\
090419$^{\star}$ & 460.7 & $(2.7 \pm 0.3) \times 10^{-6}$ & - & $(3.3 \pm 0.5) \times 10^{-9}$ & $1.3 \pm 0.2$\\
090807A & 146.4 & $(2.2 \pm 0.2) \times 10^{-6}$ & $0.7 \pm 0.2$ & $(8.6 \pm 0.9) \times 10^{-9}$ & $2.1 \pm 0.2$\\
091104 & 107.1 & $(7.6 \pm 1.2) \times 10^{-7}$ & $0.4 \pm 0.1$ & $(3.0 \pm 0.5) \times 10^{-9}$ & $1.7 \pm 0.3$\\\hline
\end{tabular}
}
\tablenotetext{$\dagger$}{Possible GRB.}
\tablenotetext{$\ddagger$}{Incomplete data to measure the fluence and peak flux.}
\tablenotetext{$\star$}{Poor statistic to measure the 1-s peak flux form the spectrum.}
\end{center}
\end{table}

We found 33 bursts (out of the 72 image triggered GRBs) with no optical counterpart observed,
which are listed in Table~\ref{tab:bat_33_IT_grb}, by looking through the Gamma-ray bursts Coordinates
Network (GCN) circulars.  These may be candidates for being 
high-redshift bursts with $z \gtrsim 6$, and might include Pop.~III GRBs.  
Their $T_{90}$ durations are relatively long, 
but not as long as the $\sim 1\;$day, predicted for Pop.~III GRBs in our model. 
These $T_{90}$ durations are just the time intervals during which BAT was able to detect 
90\% of the photons from the sources, which is not necessarily the same as the real duration 
of the bursts, although it provides a useful uniform measure approximating this quantity.
If the burst flux is marginally above the BAT threshold initially and it gradually declines, 
the $T_{90}$ duration could be much shorter than the real duration.

A bright external shock emission of a Pop.~III GRB
would trigger the BAT even if the prompt emission flux is below 
the detection threshold. We calculated the $64\;$s photon fluences of the external shock emission 
in the $15-50\;$keV band for the cases of Figures~\ref{fig:example_np} and \ref{fig:example_sp}
and obtained $S_{\rm ph} \simeq 0.33\;{\rm ph}\;{\rm cm}^{-2}$ 
and $\simeq 1.5\;{\rm ph}\;{\rm cm}^{-2}$, respectively. The photon fluence in the latter case
is marginally above the effective threshold $\sim 1\;{\rm ph}\;{\rm cm}^{-2}$.
For the case of $1+z=10$ (Figure~\ref{fig:example_z10}), we obtained 
$S_{\rm ph} \simeq 2.0\;{\rm ph}\;{\rm cm}^{-2}$ for the negligible pair production case
and $S_{\rm ph} \simeq 7.5\;{\rm ph}\;{\rm cm}^{-2}$ for the significant pair production case.
These indicate that BAT will be triggered by the external shock emission in some cases.
We have a rough relation $S_{\rm ph} \propto \epsilon_e L_{\rm iso} f(p) d_L^{-2}$ 
(see the end of Section~\ref{subsec:constraints}).
When the progenitor mass is smaller, leading to $L_{\rm iso}$ three times smaller than
the case of Figure~\ref{fig:example_np}, the external shock emission for the case of
$\epsilon_{e,-1} \simeq f(p) \simeq 1$ and $1+z \gtrsim 10$ will not trigger BAT (where 
the pair production is negligible for $n_0 < 10^4$).
In the next section, we use a specific model of the Pop.~III GRB prompt emission to examine
their detectability.

\subsection{Prompt photospheric emission}
\label{subsec:prompt}

Pop.~III GRB jets are likely to be dominated by Poynting-flux, as discussed in 
Section~\ref{sec:jet_model}.
The prompt emission mechanism of Poynting-dominated GRB jets (as opposed to the afterglow,
on which we concentrated thus far) has been actively discussed in the literature 
\citep[e.g.,][]{thompson94,meszaros97,spruit01,lyutikov06}.
The jet may have a subdominant thermal energy component of electron-positron pairs and photons,
so that the emission from the photosphere can be bright. In addition to this, above the photosphere,
the magnetic field could be directly converted into radiation via magnetic reconnection or the 
field energy could be converted into particle kinetic energy which can produce non-thermal radiation
via shocks. The existence of the latter emission components is uncertain and they are currently difficult 
to model. Thus, for simplicity we focus on the photospheric emission, which is essentially unavoidable. 
Such photospheric emission models of the prompt emission are viable also for baryonic jets, which 
could work for Pop.~I/II GRBs \citep[e.g.,][]{meszaros00,rees05,ioka07,toma10}.  MR10 developed the
Poynting-dominated jet model of \citet{meszaros97} for a Pop.~III GRB jet, and estimated the 
luminosity and temperature of the photospheric emission. Here we recalculate this emission
component, taking into account the collimation of the outflow.

Let us assume for simplicity that the opening angle of the jet is roughly constant from the base 
of the jet, $r_l = 2 g_h R_h \simeq 9.4 \times 10^7\; g_h M_{h,2.5}\;$cm where $g_h$ is a numerical 
factor, out to the external shock region. The isotropic-equivalent luminosity of the jet is 
given by $L_{\rm iso} = L_{\rm BZ} (2/\theta_j^2) \simeq 4.4 \times 10^{53}\;
(a_h^2/\alpha_{-1}\beta_1) M_{d,2.5} t_{d,4}^{-1} \theta_{j,-1}^{-2}\;{\rm erg}\;{\rm s}^{-1}$. 
Denoting by $\sigma$ the ratio of the Poynting energy flux and the particle energy flux at the 
base, the comoving temperature of the flow is estimated as
$T'_l = (L_{\rm iso}/[(1+\sigma) 4\pi r_l^2 c a \Gamma_l^2])^{1/4} \simeq 5.3 \times 10^5\;
[(1+\sigma)/10]^{-1/4} L_{53.6}^{1/4} r_{l,8}^{-1/2} \Gamma_l^{-1/2}\;{\rm eV}/k$, where 
$L_{53.6} = L_{\rm iso}/10^{53.6}\;{\rm erg}\;{\rm s}^{-1}$, $r_{l,8} = r_l/10^8\;{\rm cm}$,
and $\Gamma_l$ is the bulk Lorentz factor of the flow at the base.
The dynamics of the flow while it is optically thick
is governed by energy conservation ($L_{\rm iso} = {\rm const.}$),
entropy conservation ($r^2 {T'}^3 \Gamma = {\rm const.}$), and the MHD condition for the flow 
velocity to be close to the light speed (the lab-frame field strength $B \propto r^{-1}$). 
The last condition indicates that the Poynting energy is conserved, and so the particle energy
is also conserved, i.e., $r^2 {T'}^4 \Gamma^2 = {\rm const.}$ and $\sigma = {\rm const.}$ Then we 
have $T' \propto r^{-1}$ and $\Gamma \propto r$. At the photosphere radius $r = r_a$ where the 
electron-positron pairs recombine, the temperature is given by $kT'_a \sim 17\;$keV. This leads to
\begin{eqnarray}
r_a &\simeq& 3.1 \times 10^9\; [(1+\sigma)/10]^{-1/4} L_{53.6}^{1/4} r_{l,8}^{1/2} \Gamma_l^{-1/2}\; 
{\rm cm}, \\
\Gamma_a &\simeq& 31\; [(1+\sigma)/10]^{-1/4} L_{53.6}^{1/4} r_{l,8}^{-1/2} \Gamma_l^{1/2}.
\end{eqnarray}
The observed temperature and the bolometric energy flux of the photospheric emission are then
\begin{eqnarray}
kT_a &\simeq& \frac{\Gamma_a kT'_a}{1+z} \nonumber \\
&\simeq& 26\; [(1+\sigma)/10]^{-1/4} L_{53.6}^{1/4} 
r_{l,8}^{-1/2} \Gamma_l^{1/2} [(1+z)/20]^{-1}\;{\rm keV}, \\ 
F_a &\simeq& \frac{L_{\rm iso}/(1+\sigma)}{4\pi d_L^2} \nonumber \\
&\simeq&
7.1 \times 10^{-9}\;[(1+\sigma)/10]^{-1} L_{53.6} d_{L,20}^{-2} \;{\rm erg}\;{\rm cm}^{-2}\;{\rm s}^{-1}.
\end{eqnarray}
The photospheric photons can be scattered by MHD turbulence or Alfv\'{e}n waves, induced by e.g.,
the interaction of the jet with the stellar envelope, into a power-law spectrum extending up to
comoving photon energies $\sim m_e c^2$ \citep{thompson94}\footnote{
If the power-law spectrum extends to energies much higher than $\sim m_e c^2$, e.g., due to
magnetic dissipation, as argued in MR10, copious pair formation would ensue, which would form a
new (pair) photosphere at a larger radius \citep[e.g.,][]{rees05}.}.
The emission from the photosphere of Pop.~III GRB would thus have
a black-body peak, and may have a non-thermal tail extending (in the observer frame) to
photon energies $\varepsilon \sim \Gamma_a m_e c^2 /(1+z) \sim 1\;$MeV.
This photospheric emission may be detected until $t_{\rm obs} = t_{d,{\rm obs}}$. We plot this in 
Figures~\ref{fig:example_np}, \ref{fig:example_sp}, and \ref{fig:example_z10} with dotted lines for the 
case of $1+\sigma = 10$ and $L_{53.6} = r_{l,8} = \Gamma_l = 1$.

We calculated the photon fluences in $64\;$s in the $15-50\;$keV band for the above parameter sets
as $S_{\rm ph} \simeq 1.4\;{\rm ph}\;{\rm cm}^{-2}$ for the case of $1+z=20$ and
$\simeq 1.3\;{\rm ph}\;{\rm cm}^{-2}$ for the case of $1+z=10$.
Thus this emission can be marginally detected by the image trigger of BAT.  
The value of $\sigma$ is highly uncertain, similar to $R_*$ and
$\theta_j$ for a specific value of $M_*$.
For $L_{\rm iso}/(1+\sigma)$ three times smaller than the above case (and similar values of 
$r_l$ and $\Gamma_l$), the photon fluence is calculated as $< 1\;{\rm ph}\;{\rm cm}^{-2}$ 
both for $1+z = 20$ and $1+z = 10$, and thus BAT is not expected to be triggered.

\subsection{Pop.~III GRB Rate}

The Pop.~III GRB rate is largely uncertain, and has only been inferred from theoretical models.
The observed rate of Pop.~III GRBs originating between redshifts $z$ and $z+dz$ is computed by
\begin{equation}
\frac{d\dot{N}_{\rm GRB}^{\rm obs}}{dz} = \psi_*(z) \eta_{\rm GRB}(z) P(z) \frac{1}{1+z} \frac{dV}{dz}
\end{equation}
where $\psi_*(z)$ is the Pop.~III star formation rate (SFR) per unit comoving volume, 
$\eta_{\rm GRB}(z)$ is the efficiency of the GRB formation, $P(z)$ is the detection efficiency, 
i.e., the ratio of the Pop.~III GRBs which would be detected by a specific instrument out of 
the entire number of Pop.~III GRBs, and $dV/dz$ is the comoving volume element of the observed area 
per unit redshift.  The additional factor $1/(1+z)$ represents
the cosmological time dilation effect. 

The factors $\psi_*(z)$ and $\eta_{\rm GRB}(z)$ are both highly uncertain for the Pop.~III VMSs
as well as for the Pop.~I/II stars.  
Just for a concrete discussion, here we use $\psi_*(z)$ predicted
by using the extended Press-Schechter formalism and the current observational results on the 
SFR for $z \lesssim 6$ \citep{bromm06} \citep[see also][]{naoz07}. 
For Pop.~I/II stars, they assumed a $z$-independent
$\eta_{\rm GRB} \simeq 2 \times 10^{-9}\;M_{\odot}^{-1}$ to set the total Pop.~I/II GRB rate 
observed by {\it Swift} BAT to be $\sim 90\;{\rm yr}^{-1}$. (Note that they
assumed that {\it Swift} BAT covers $4\pi$ of the sky, while it actually covers only $\sim 2\pi/3$
of the sky. Thus the GRB formation efficiency should be normalized as $\sim 6$ times their
adopted value, $\eta_{\rm GRB} \simeq 1.2 \times 10^{-8}\;M_{\odot}^{-1}$.)
Their result predicts the Pop.~I/II GRB rate observed by BAT at $z > 6$ to be $\sim 10\;{\rm yr}^{-1}$.
Since a small fraction $\sim 25\%$ of GRBs detected by BAT have redshifts determined, because of 
bad conditions for optical and near-IR observations \citep[cf.][see also our implication from the 
image triggered GRBs in Section~\ref{subsec:BAT} and in Table~\ref{tab:bat_33_IT_grb}]{fynbo09}, 
the predicted rate of Pop.~I/II GRBs with $z$ determined would be $\sim 3\;{\rm yr}^{-1}$.
This is somewhat higher than the current observed rate, $0.6\;{\rm yr}^{-1}$, 
i.e., 3 GRBs (GRB 050904, GRB 080913, and GRB 090423) during the 5-yr operation of {\it Swift}, 
but their model of $\psi_*(z)$ and $\eta_{\rm GRB}$ is not interpreted as unacceptable, 
taking into account the uncertainties of the theoretical calculations and the poor statistics of the 
current observed data.
For Pop.~III stars, they assumed the same $z$-independent $\eta_{\rm GRB}$ as Pop.~I/II stars
and computed the nominal Pop.~III GRB rate observed by BAT to be 
$R_{\rm BAT} \approx z d\dot{N}_{\rm GRB}^{\rm obs}/dz \sim 0.03\;{\rm yr}^{-1}$ 
for bursts around $z \sim 20$ and $\sim 0.3\;{\rm yr}^{-1}$ for bursts around $z \sim 10$.  

The detection efficiency $P(z)$ is computed by assuming the GRB luminosity function and the 
detection threshold. \citet{bromm06} assumed that the Pop.~III GRB luminosity function is the 
same as that of Pop.~I/II GRBs and took a BAT detection threshold of $f_{\rm ph,lim} \simeq 0.2\;
{\rm ph}\;{\rm cm}^{-2}\;{\rm s}^{-1}$.  We can assume, however, that the Pop.~III GRBs have
a different luminosity function, with a brighter membership than Pop.~I/II GRBs, and take 
the effective BAT threshold deduced by our analysis of the image trigger bursts 
(see Section~\ref{subsec:BAT}), in which case $R_{\rm BAT}$ can be larger than the above estimate.

In order to estimate $R_{\rm BAT}$ in our Pop.~III GRB model, let us first presume the 
observed rate (for $\sim 2\pi/3$ of the sky) if we would detect the entirety of the 
Pop.~III GRBs originating around $z \sim 10-20$ without considering the detection efficiency, $R_w$,
based on the calculations of \cite{bromm06}. The detection efficiency is computed by
\begin{equation}
P(z) = \int^{\infty}_{L_{\rm ph,lim}(z)} p(L_{\rm ph}) dL_{\rm ph},
\end{equation}
where $L_{\rm ph}$ is the isotropic-equivalent photon luminosity of a burst, and $p(L_{\rm ph})$ is the 
luminosity function normalized as $\int^{\infty}_0 p(L_{\rm ph}) dL_{\rm ph} = 1$. 
They adopted $L_{\rm ph,lim} = 4\pi d_L^2 f_{\rm ph,lim} \simeq 1 \times 10^{60}\;{\rm ph}\;{\rm s}^{-1}$ 
for bursts around $z \sim 20$ and $\simeq 2 \times 10^{59}\;{\rm ph}\;{\rm s}^{-1}$ for bursts
around $z \sim 10$. The detection efficiency is then roughly estimated to be 
$\sim p(L_{\rm ph,lim}) L_{\rm ph,lim}/ p(L_p) L_p \sim 0.2$ for $z \sim 20$ and $\sim 0.5$ for $z \sim 10$, 
where $L_p$ provides the peak of the function of $p(L_{\rm ph}) L_{\rm ph}$.
Thus, we obtain $R_w \sim R_{\rm BAT}/0.2 \sim 0.2\;{\rm yr}^{-1}$ for $z \sim 20$ and 
$\sim R_{\rm BAT}/0.5 \sim 0.6\;{\rm yr}^{-1}$ for $z \sim 10$.

According to our study in Sections~\ref{subsec:BAT} and \ref{subsec:prompt}, 
the photospheric prompt emission and/or the 
external shock emission of Pop.~III GRBs can be detected by the BAT image trigger for
our fiducial set of parameters, $M_{d,2.5} \simeq M_{h,2.5} \simeq R_{*,12} \simeq \theta_{j,-1}
\simeq \epsilon_{e,-1} \simeq f(p) \simeq (1+\sigma)/10 \simeq 1$ and $1+z \gtrsim 10$,
while not detected for $L_{\rm iso}$ just three times smaller than that for the above set
of parameters. Thus, if the distribution of the parameters of the Pop.~III VMSs clusters
around our fiducial set of parameters, BAT would detect, say, about half of 
the Pop.~III GRBs out of the whole Pop.~III GRBs, i.e., the detection rate could be
$R_{\rm BAT} \sim R_w/2 \sim 0.1\;{\rm yr}^{-1}$ for $z \sim 20$ and 
$\sim 0.3\;{\rm yr}^{-1}$ for $z \sim 10$. 
These values have large uncertainties, but imply that the 5-yr operation of {\it Swift} so far may 
already have detected a Pop.~III GRB, or may detect it in the near future, if the factor $\psi_*(z)   
\times \eta_{\rm GRB}$ is given as above.

\section{Late-Time Radio Afterglows}
\label{sec:radio}

We have focused so far on the high-energy emission just before and near the beginning of the 
external shock self-similar phase ($t_{\rm obs} \simeq t_{d,{\rm obs}}$, see Figure~\ref{fig:lightcurve})
to constrain the physical parameters $E_{\rm iso}$ and $n$ of 
Pop.~III GRBs, and to examine their detectability by BAT. In this section, we argue that 
the radio afterglows of Pop.~III GRBs in the self-similar phase ($t_{\rm obs} > t_{d,{\rm obs}}$)
can be so bright that they also provide powerful tools for constraining the event rate.  
At $t_{d,{\rm obs}}$, the flux of the external shock emission shown in Figure~\ref{fig:example_np} at 
$\nu_a \simeq 230\;E_{57.6}^{1/6} t_{d,4}^{-1/2} n_0^{1/6} [(1+Y)/3.7]^{-1/3} [(1+z)/20]^{-1}\;$GHz 
is given by
\begin{eqnarray}
F_{\nu_a} \simeq
2.6\; \epsilon_{B,-2}^{-1/4} E_{57.6}^{2/3} n_0^{-1/12} [(1+Y)/3.7]^{-5/6}
[(1+z)/20] d_{L,20}^{-2}\;{\rm Jy}.
\end{eqnarray}
The break frequency $\nu_a$ decreases (as $t_{\rm obs}^{-1/2}$) but the flux at $\nu_a$ stays constant 
after $t_{d,{\rm obs}}$, at least 
until the epoch when the jet effects become significant. 
{\it This indicates that the Pop.~III GRB afterglows can be very bright radio sources, 
despite their large distances.}

We briefly compute the light curves at various frequencies in the radio bands, 
100~GHz, 5~GHz, 1~GHz, and 70~MHz, in our fiducial case,
$E_{57.6} = t_{d,4} = n_0 = \epsilon_{B,-2} = \epsilon_{e,-1} = f(p) = \theta_{j,-1} = 1$.
As discussed above, the temporal evolution of the characteristic quantities of the external
shock in the self-similar expansion phase can be obtained by replacing $t_d$ by the variable $t_{\rm obs}$ 
and taking the other parameters as constant in the equations of the general afterglow model 
(shown in Appendix). 
The jet effects are significant when the Lorentz factor of the shocked fluid is 
$\Gamma \simeq \theta_j^{-1}$, i.e., at the observer's time
\begin{eqnarray}
t_{\theta,{\rm obs}} \simeq
\left(\frac{E_{\rm iso} \theta_j^8}{4\pi n m_p c^5}\right)^{1/3} (1+z) 
\simeq 9.9\times10^2 \; E_{57.6}^{1/3} n_0^{-1/3} \theta_{j,-1}^{8/3} 
[(1+z)/20]\;{\rm day}.
\end{eqnarray}
After $t_{\theta,{\rm obs}}$, the temporal evolution of the characteristic quantities is obtained by
replacing $E_{\rm iso}$ by $t_{\rm obs}^{-1}$ and $t_d$ by $t_{\rm obs}$, respectively, and taking the other
parameters as constant \citep{sari99}. Such an evolution is derived under the assumption that
the shocked fluid expands sideways rapidly after $t_{\theta,{\rm obs}}$. Recent detailed hydrodynamic 
simulations \citep{wzhang09,granot07} have shown that the shocked fluid only undergoes a slow sideways
expansion, while the afterglow light curves can be still approximated
by those predicted by \citet{sari99} especially in the radio band. 
(In the optical and X-ray bands, the spectrum is so soft that the limb-brightening effect is 
significant. Then the brightest portion at an angle $\theta \sim \Gamma^{-1}$ from the line of sight 
becomes missing at $t_{\theta,{\rm obs}}$, which causes a steeper light curve than $t_{\rm obs}^{-p}$.) 

In an analytical treatment \citep[e.g.,][]{sari99} the rapid sideways expansion of the fluid
would lead to a nearly spherical ($\theta_j \sim 1$) configuration of the shocked fluid
leading to the non-relativistic Sedov-von Neumann-Taylor (SNT) solution, starting around 
observer's time $t_{\rm SNT,obs} \simeq 
[E_{\rm iso} \theta_j^2/(4\pi n m_p c^5)]^{1/3} (1+z)
\simeq 9.9\times 10^4\;E_{57.6}^{1/3} n_0^{-1/3} \theta_{j,-1}^{2/3} [(1+z)/20]\;$day. 
However, the detailed hydrodynamic simulations \citep{wzhang09} show that the much slower 
sideways expansion results in the later start of the SNT phase. The starting time is shown to be 
a few times $t_{\rm NR,obs}$
\begin{eqnarray}
t_{\rm NR,obs} \simeq
\left(\frac{E_{\rm iso}}{4\pi n m_p c^5}\right)^{1/3} (1+z) 
\simeq 4.6 \times 10^5\;E_{57.6}^{1/3} n_0^{-1/3} [(1+z)/20]\;{\rm day},
\end{eqnarray}
where the shocked fluid is still highly collimated. 
The temporal evolution of the characteristic quantities in the SNT phase
is obtained by using $r \propto t_{\rm obs}^{2/5}$ and $v \propto t_{\rm obs}^{-3/5}$, 
where $v$ is the velocity of the shocked fluid, and the internal energy density is $\propto v^2$.
This results in a light curve with a shallower decay at $\nu > \nu_a$ (or a steeper rise at $\nu < \nu_a$) 
in the SNT phase, compared to the decay in the prior phase. Therefore, the assumption of the slow 
sideways expansion and the late SNT phase based on the numerical simulations leads to smaller number 
of off-axis observers (due to the high collimation) and dimmer radio fluxes than that of 
the rapid sideways expansion based on the approximate analytical arguments. 
We here take the former, conservative assumption of the slow sideways expansion.

\begin{figure}
\epsscale{0.75}
\plotone{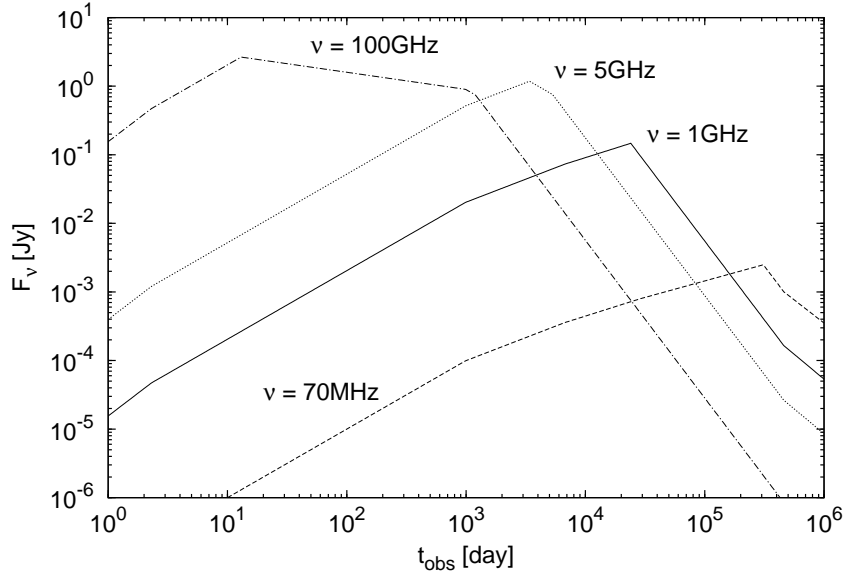}
\caption{
Radio light curves at frequencies, 100~GHz (dot-dashed line), 5~GHz (dotted line), 1~GHz (solid line), 
and 70~MHz (dashed line), of a Pop.~III GRB
at $1+z = 20$ with typical parameters $E_{57.6} = t_{d,4} = n_0 = \epsilon_{B,-2} = 
\epsilon_{e,-1} = f(p) = \theta_{j,-1} = 1$. The important observer's times are the jet duration 
$t_{d,{\rm obs}} \simeq 2.3\;$day, the jet break time $t_{\theta,{\rm obs}} \simeq 9.9 \times
10^2\;$day, and the time when the shock becomes non-relativistic $t_{\rm NR,obs} \simeq 4.6\times
10^5\;$day.
}
\label{fig:radiolc}
\end{figure}

For calculating the radio emission we need to compute only the synchrotron emission of the 
original electrons. At late times, the number of pairs is typically small, and the pair emission 
is negligible. The SSC emission is not relevant in the radio bands.  The results are plotted
in Figure~\ref{fig:radiolc}. Here we confirmed the assumptions that all the electrons in the shocked 
region remain relativistic at least until $t_{\rm obs} = 10^6\;$day.

Figure~\ref{fig:radiolc} shows that the radio afterglows of Pop.~III GRBs can be very bright
with a very long duration. These could have been detected as quasi-steady point sources by the 
radio survey observations. As far as we know, the current largest radio survey data is based on
the Very Large Array (VLA) FIRST survey \citep{white97}, which observed a large area mainly around 
the north Galactic cap at 1.4~GHz, covering $\sim 1/5$ of all the sky.\footnote{\citet{levinson02}
and \citet{galyam06} did not find any radio transient sources like GRB afterglows with 
timescales of significant flux changes $\sim 5\;{\rm yr}$ by comparison between the NVSS (spanned over 
1993-1996) and FIRST (1994-2001) catalogs, which effectively cover $\sim 1/17$ of the sky. 
This indicates that $R_{w,4\pi} \lesssim 17/5 \sim 3\;{\rm yr}^{-1}$.}
For the threshold, $\sim 6\;$mJy at 1.4~GHz, a Pop.~III GRB radio afterglow at $1+z=20$ (at $1+z=10$) 
can be observed for $t_{rd,{\rm obs}} \sim 300\;$yr (for $t_{rd,{\rm obs}} \sim 200\;$yr). 
Therefore, if we denote by $R_{w,4\pi}$ the all-sky Pop.~III 
GRB rate in unit of ${\rm yr}^{-1}$, the number of the Pop.~III radio afterglows that would have 
been detected in that survey is estimated to be 
$\sim (R_{w,4\pi}/5) t_{rd,{\rm obs}} \sim 60\;(R_{w,4\pi}/1\;{\rm yr}^{-1}) 
(t_{rd,{\rm obs}}/300\;{\rm yr})$ for bursts at $1+z \sim 20$, and
$\sim 200\;(R_{w,4\pi}/4\;{\rm yr}^{-1}) (t_{rd,{\rm obs}}/200\;{\rm yr})$ for bursts at $1+z \sim 10$.
A detailed analysis of the FIRST data would thus provide a powerful constraint on $R_{w,4\pi}$ of
Pop.~III GRBs such as discussed in this paper (even no sources like our model calculations would 
provide an upper limit on the rate).  The Pop.~III radio sources could have X-ray counterparts.
In our model the late time X-ray afterglow is dominated by the SSC component, whose flux in the
$0.3-10\;$keV range can be $> 10^{-15}\;{\rm erg}\;{\rm cm}^{-2}\;{\rm s}^{-1}$ for 
$t_{\rm obs} \lesssim 100\;$yr (which can be detected by Chandra X-ray Observatory).

The above predicted numbers of the Pop.~III GRB radio afterglows are just based on the values
of $t_{rd,{\rm obs}}$ calculated for our fiducial set of parameters. In order to obtain more
realistic numbers taking into account the distributions of the model parameters, we require
to calculate the radio light curves for large ranges of parameter values. However, the radio
light curves may highly depend on the parameters $E_{\rm iso}, t_d, n, \epsilon_B, \epsilon_e$, 
and $f(p)$. In other words, there are many patterns of light curves depending on the orders of
characteristic times $t_{a=c}, t_{m=a},$ and $t_{m=c}$, which are the times when $\nu_a$ crosses
$\nu_c$, when $\nu_m$ crosses $\nu_a$, and $\nu_m$ crosses $\nu_c$, respectively, as well as
$t_d, t_{\theta},$ and $t_{\rm NR}$ (we have $t_d < t_\theta < t_{m=a} < t_{m=c} < t_{a=c} < t_{\rm NR}$
for our adopted parameter set). A thorough
model analysis with different parameter sets would deserve another future work.

Recent radio transient searches without primary detections at any other frequencies have begun
to open new observational frontiers \citep[e.g.,][]{niinuma07,bower07,ofek10}, but currently they
do not appear to have good potentials for detecting Pop.~III GRB radio afterglows.
The interferometric drift-scanning observation with the Waseda Nasu Pulsar Observatory in Japan may 
scan a large area at $1.4\;$GHz \citep{niinuma07}, but its detection thereshold, $\sim 0.3\;$Jy, 
is not sufficiently high for searching our typical Pop.~III radio afterglows. \citet{bower07}
reported results of a survey for transient sources by using archival data obtained with VLA observation
of a single field at 5 or 8.4~GHz, but the field is very small, $< 0.1\;{\rm deg}^2$.

The absorption line at 21~cm (associated with transitions between the hyperfine levels of the 
hydrogen atom) seen in the continuum radio spectrum of high-redshift sources would be one of the 
promising tools to reveal the reionization histroy of the universe \citep{tozzi00,shaver99}.
GRB radio afterglows have been considered as candidate backlighting sources, but the bursts with 
the usual or slightly higher total energies have been found to be too dim for this aim 
\citep{furlanetto02,ioka05}. However, the high-redshift GRBs arising from Pop.~III VMSs, 
such as discussed in this paper, can emit a sufficiently bright radio afterglow to be of
interest. For our fiducial parameter values, the radio flux at $\nu = 1420\;{\rm MHz}/
(1+z)$ is estimated to be $\simeq 2.4\;$mJy (with peak time $t_{p,{\rm obs}} \simeq 850\;$yr) for 
$1+z=20$ and $\simeq 6.4\;$mJy (with peak time $t_{p,{\rm obs}} \simeq 430\;$yr) for $1+z=10$, 
which are comparable to the lower bound on the radio flux for detection of the 21~cm 
absorption line, $\sim 1-10\;$mJy \citep{ioka05}. Thus the 21~cm absorption line could be 
marginally detected in the Pop.~III GRB radio afterglows.  A radio survey with telescopes like
Low Frequency Array (LOFAR)\footnote{http://www.lofar.org.} 
could detect such radio afterglows and determine their redshifts by
their 21~cm lines themselves.  The predicted number of detections is 
$\sim R_{w,4\pi} (\Omega/4\pi) t_{p,{\rm obs}}
\sim 1\;(R_{w,4\pi}/1\;{\rm yr}^{-1}) (\Omega/50\;{\rm deg}^2)
(t_{p,{\rm obs}}/850\;{\rm yr})$ for bursts at $1+z \sim 20$, and
$\sim 1\;(R_{w,4\pi}/4\;{\rm yr}^{-1}) (\Omega/20\;{\rm deg}^2)
(t_{p,{\rm obs}}/430\;{\rm yr})$ for bursts at $1+z \sim 10$, 
where $\Omega$ is the solid angle of the survey area.
Note that the jets will keep collimated even around $\sim t_{\rm NR,obs}$ due to the slow
sideways expansions \citep{wzhang09}, so that off-axis viewings of the afterglows are too 
dim to detect.\footnote{If the jets were instead to undergo a rapid sideways expansion,
as suggested by the approximate analytical treatment of the transition to the non-relativistic
regime \citep[e.g.,][]{sari99}, the peak fluxes and peak times are slightly larger than 
the above estimates, i.e.,
$F_p \simeq 4.3\;$mJy with $t_{p,{\rm obs}} \simeq 1300\;$yr at $1+z=20$,
and $F_p \simeq 11\;$mJy with $t_{p,{\rm obs}} \simeq 630\;$yr at $1+z=10$. 
Since $t_{p,{\rm obs}} > t_{\rm SNT,obs}$ and thus the fluids would be nearly spherical at the peak
time, the off-axis observers can detect the emission. Therefore the predicted number of detections 
would be much larger, by a factor of $\theta_j^{-2} = 10^2 \theta_{j,-1}^{-2}$.}

\section{Summary}
\label{sec:discussion}

Pop.~III GRBs may have an isotropic-equivalent energy $E_{\rm iso} \gtrsim 10^{57}\;$erg and 
a cosmological-rest-frame duration $t_d \gtrsim 10^4\;$s. We have calculated the external shock 
emission spectrum at $t_{d,{\rm obs}}$ in detail based on the standard model 
\citep{mesAG97,sari98,sari01,nakar09}. This model can explain many of the X-ray/optical/radio 
afterglows detected so far \citep[e.g.,][]{panaitescu02,liang07}
as well as the very early high-energy afterglows detected by {\it Fermi} LAT 
\citep[e.g.,][]{kumar09,depasquale10,corsi10}.

We found that the external shock emission at $t_{d,{\rm obs}}$ can be detected by 
{\it Fermi} LAT and {\it Swift} XRT/BAT, whose flux leads to
a constraint on $E_{\rm iso}$ by using the source redshift (and distance) that will be 
determined by the observation of the Ly$\alpha$ drop-off in the IR band.  The detection of a burst at 
$z \gtrsim 10$ with $E_{\rm iso} \gtrsim 10^{57}\;$erg and $t_d \gtrsim 10^4\;$s would be a very 
strong indication that this is a GRB arising from a Pop.~III VMS with $M_* \gtrsim 300\;M_\odot$. 
This indication should be complemented with the constraint on the metal abundances in the surrounding
medium through high resolution IR and X-ray spectroscopy.

One of the important findings of the present study is that the $\gamma\gamma$ self-absorption 
break at energy $\varepsilon_{\gamma\gamma}$ in some cases of the external shock emission spectrum of a 
Pop.~III GRB is expected to be observable in the LAT energy range.  Given that the prompt emission 
at $t_{d,{\rm obs}}$ does not hide the external shock emission in the LAT range and 
$\varepsilon_{\gamma\gamma}$
is well below the values of the EBL $\gamma\gamma$ cutoff energy expected for practically all EBL models
with the determined source redshift,
we have shown that the flux and the energy $\varepsilon_{\gamma\gamma}$, together 
with {\it Swift} XRT data, can lead to a constraint on the value of the external density $n$.
The constraint on $n$ would provide invaluable information about the environment and the radiative 
feedback processes of Pop.~III stars. 
 
Putting constraints on $n$ from the $\gamma\gamma$ self-absorption break is a fairly new 
method, while constraints on $n$ from the multi-wavelength observations of the late-time 
afterglows of Pop.~I GRBs are common \citep[e.g.,][]{panaitescu02,wijers99}.  Here 
we have discussed the conditions under which these methods can be used in Pop. III GRBs. 
One caveat is that we have assumed that the electron acceleration processes works 
uniformly in the emitting region for calculating the afterglow spectra.  
As discussed below Equation~(\ref{eq:condition_p}) in Appendix,
it is also possible that the 
electron acceleration works only near the shock front, where $\varepsilon_{\gamma\gamma}$ may 
be far above the LAT energy range, and $\varepsilon_M$ depends on the unknown upstream magnetic
field strength, which do not allow us to constrain $n$.

In order to identify the direction to GRB afterglows on the sky for observations with {\it Swift}, 
{\it Fermi}, and the IR telescopes, the emission needs to be high enough to trigger the large 
field instrument BAT.  A reasonable trigger
threshold of {\it Swift} BAT can be estimated by focusing on the image trigger
mode, since it is based on a criterion with a fixed time-scale and energy band, while more
general BAT threshold including the usual rate trigger is too complicated to estimate.
The image trigger mode may be suitable for detecting weak and less-variable bursts like 
very-high-redshift bursts. We have used the samples in the BAT2 catalog and deduced the detection 
threshold of the BAT image trigger to be 
$\sim 1\;{\rm ph}\;{\rm cm}^{-2}$ for the $64\;$s interval in the $15-50\;$keV band. 
We have calculated the prompt photospheric emission flux of Pop.~III GRBs with typical parameters,
and shown that these can be marginally detected by BAT. The external shock emission can also 
trigger BAT without the prompt emission trigger in some cases.

We have also briefly shown that the Pop.~III GRB late-time radio afterglows can be very bright.
For our fiducial parameters, the peak flux at $1\;$GHz is $\simeq 140\;$mJy, which could be 
identified in the VLA FIRST survey data. This survey covered a sufficiently large area, which
would provide a powerful upper bound on the rate of the Pop.~III GRBs.
The peak flux of the late-time $70\;$MHz radio afterglow for our fiducial parameters 
is $\simeq 2.4\;$mJy, in which 21~cm absorption lines could be detected. This would provide 
a measure of the neutral hydrogen fraction in the IGM around the Pop.~III star. 

Putting constraints on the properties of Pop.~III stars has recently become of great importance 
in modern cosmology. Planned IR surveys will be able to probe Pop.~III stars. However, it is 
difficult to distinguish between a single Pop.~III VMS and a cluster of less massive 
Pop.~III stars. Thus, the detection of GRBs with very high $E_{\rm iso}$ and very long $t_d$ 
could provide critical, `smoking gun' evidence for the existence of VMSs. Multi-wavelength 
observations of such GRBs with {\it Swift}, {\it Fermi}, and ground-based IR and radio telescopes 
should provide us with invaluable information on Pop.~III stars and their environments.

\acknowledgements
We thank D.~N.~Burrows, A.~D.~Falcone, D.~B.~Fox, A.~Gal-Yam,
S.~Gao, K.~Murase, S.~Naoz, and the anonymous referee for useful comments.
We acknowledge NASA NNX09AT72G, NASA NNX08AL40G, and NSF PHY-0757155 for partial support.
PM is grateful for the hospitality of Fermilab and the Institute of Astronomy, Cambridge
University, during part of this project.

\appendix

\section{General Model}
\label{subsec:ag_model}

The afterglow emission spectrum at the time $t_d$ is determined by the radius $r_d$ and 
the Lorentz factor $\Gamma_d$ of the shocked fluid at this time, which are given via the relations 
$E_{\rm iso} \simeq 4\pi r_d^3 \Gamma_d^2 n m_p c^2$ and $r_d \simeq c \Gamma_d^2 t_d$. 
Here we have assumed for simplicity that the circumburst medium density is uniform, 
$n = 1\;n_0\;{\rm cm}^{-3}$. These two equations lead to
\begin{eqnarray}
r_d &\simeq& \left(\frac{E_{\rm iso} t_d}{4\pi n m_p c}\right)^{1/4} 
\simeq 2.8 \times 10^{18}\; E_{57.6}^{1/4} t_{d,4}^{1/4} n_0^{-1/4} \;{\rm cm}\\
\Gamma_d &\simeq& \left(\frac{E_{\rm iso}}{4\pi n m_p c^5 t_d^3}\right)^{1/8}
\simeq 97\; E_{57.6}^{1/8} t_{d,4}^{-3/8} n_0^{-1/8},
\end{eqnarray}
where $E_{57.6} = E_{\rm iso}/10^{57.6}\;{\rm erg}$ and $t_{d,4} = t_d/10^4\;{\rm s}$.
The magnetic field strength in the shocked region scales as
\begin{equation}
B \simeq (32\pi \epsilon_B n m_p c^2)^{1/2} \Gamma_d \simeq 3.8\; \epsilon_{B,-2}^{1/2}
E_{57.6}^{1/8} t_{d,4}^{-3/8} n_0^{3/8}\;{\rm G},
\end{equation}
where $\epsilon_{B,-2} = \epsilon_B/10^{-2}$. The minimum injected electron Lorentz 
factor is 
\begin{equation}
\gamma_m \simeq \epsilon_e \Gamma_d \frac{m_p}{m_e} \frac{p-2}{p-1} 
\simeq 4.1 \times 10^3\; \epsilon_{e,-1} E_{57.6}^{1/8} t_{d,4}^{-3/8} n_0^{-1/8} f(p),
\end{equation}
where $\epsilon_{e,-1} = \epsilon_e/10^{-1}$ and $f(p) = (13/3)(p-2)/(p-1)$. 
We have assumed that all the electrons are accelerated to a power-law spectrum 
$dn/d\gamma_e \propto \gamma_e^{-p}$ for $\gamma_e \geq \gamma_m$.

The accelerated  electrons will lead to synchrotron and SSC emission. The radiative cooling 
timescale in the comoving frame is $t'_c = 6\pi m_e c/[\sigma_T \gamma_e B^2 (1+Y(\gamma_c))]$, 
where $Y(\gamma_c)$ is the luminosity ratio of the SSC to synchrotron emission for electrons 
with $\gamma_c$, while the comoving dynamical timescale is $t'_d = \Gamma_d t_d$. Thus the 
electron Lorentz factor above which the radiative cooling is more significant than the 
adiabatic cooling is 
\begin{equation}
\gamma_c \simeq \frac{6\pi m_e c}{\sigma_T B^2 \Gamma_d t_d [1+Y(\gamma_c)]}
\simeq 56\; [1+Y(\gamma_c)]^{-1} \epsilon_{B,-2}^{-1} E_{57.6}^{-3/8} t_{d,4}^{1/8} n_0^{-5/8}.
\end{equation}
All the injected electrons are radiatively cooled within the dynamical timescale if 
$\gamma_c < \gamma_m$, that is if
\begin{equation}
\epsilon_{B,-2} n_0^{1/2} [1+Y(\gamma_c)] > 1.4 \times 10^{-2}\; \epsilon_{e,-1}^{-1} E_{57.6}^{-1/2} 
t_{d,4}^{1/2} {f(p)}^{-1}.
\label{eq:condition_fc}
\end{equation}
This condition can be rewritten using Eqs.~(\ref{eq:t_d}) and (\ref{eq:E_iso}) as
\begin{equation}
M_{d,2.5}^{1/2} M_{h,2.5}^{1/4} R_{*,12}^{-3/4} n_0^{1/2} \theta_{j,-1}^{-1} a_h \beta_1^{-1/2}
(1-\epsilon_\gamma)^{1/2} > 10^{-2} \epsilon_{B,-2}^{-1} \epsilon_{e,-1}^{-1} {f(p)}^{-1} [1+Y(\gamma_c)]^{-1}.
\end{equation}
Since the masses $M_d$ and $M_h$ are expected to be positively correlated with the radius $R_*$, 
this condition is found to be satisfied for reasonable parameter ranges. We thus focus on 
the fast-cooling regime, $\gamma_c < \gamma_m$ for the emission at $t_d$.

The peak energies of the $\varepsilon F_{\varepsilon}$ spectra of the synchrotron and SSC emission 
are given by 
\begin{eqnarray}
\varepsilon_m &\simeq& \frac{3heB}{4\pi m_e c}\gamma_m^2 \frac{\Gamma_d}{1+z}
\simeq 5.4\; \epsilon_{e,-1}^2 \epsilon_{B,-2}^{1/2} E_{57.6}^{1/2} t_{d,4}^{-3/2} {f(p)}^2 
[(1+z)/20]^{-1}\;{\rm eV}, \\
\varepsilon_m^{\rm SC} &\simeq& 2\gamma_m^2 \varepsilon_m
\simeq 1.8\times10^2\; \epsilon_{e,-1}^4 \epsilon_{B,-2}^{1/2} E_{57.6}^{3/4} t_{d,4}^{-9/4} n_0^{-1/4}
{f(p)}^4 [(1+z)/20]^{-1} \;{\rm MeV},
\end{eqnarray}
respectively. If $\varepsilon_m^{\rm SC} < \Gamma_d \gamma_m m_e c^2/(1+z)$, the 
Klein-Nishina (KN) suppression of the SSC emission is not significant \citep[e.g.,][]{nakar09}.
This condition is rewritten as $\epsilon_{B,-2} < 3.2 \times 10^3 
\epsilon_{e,-1}^{-6} E_{57.6}^{-1} t_{d,4}^3 {f(p)}^{-6}$, which is satisfied for 
reasonable parameter ranges.  In the case of negligible KN effects, $Y$ does not 
depend on $\gamma_e$ for $\gamma_e \lesssim \gamma_m$, which is calculated from
\begin{equation}
Y \simeq \sigma_T \frac{r_d}{\Gamma_d} \int_{\gamma_c}^{\infty} d\gamma_e
\frac{dn}{d\gamma_e} \gamma_e^2 \simeq \tau \gamma_c \gamma_m 
\frac{p-1}{p-2} \simeq \frac{\epsilon_e}{\epsilon_B} \frac{1}{1+Y},
\label{eq:y_1}
\end{equation}
where $\tau = \sigma_T r_d n$ is the optical depth for the electron scattering.  We obtain 
$Y \approx \sqrt{\epsilon_e/\epsilon_B} \sim 3\;\epsilon_{e,-1}^{1/2} \epsilon_{B,-2}^{-1/2}$
for the case of $\epsilon_e \gtrsim \epsilon_B$, while otherwise $Y \approx \epsilon_e/\epsilon_B 
< 1$.  The SSC spectrum above $\varepsilon_m^{\rm SC}$ has a softening break at
\begin{equation}
\varepsilon_{\rm KN} \simeq \frac{\Gamma_d^2 m_e^2 c^4}{(1+z)^2 \varepsilon_m}
\simeq 1.2\; \epsilon_{e,-1}^{-2} \epsilon_{B,-2}^{-1/2} E_{57.6}^{-1/4} t_{d,4}^{3/4}
n_0^{-1/4} {f(p)}^{-2} [(1+z)/20]^{-1}\; {\rm TeV}.
\end{equation}
The fluxes at $\varepsilon_m$ and at $\varepsilon_m^{\rm SC}$ are given by 
\begin{eqnarray}
\varepsilon_m F_{\varepsilon_m} &\simeq& \frac{\epsilon_e E_{\rm iso} t_d^{-1}}{4\pi d_L^2}
\frac{p-2}{p-1} \frac{1}{1+Y} \nonumber \\
&\simeq& 1.6\times 10^{-9}\;(1+Y)^{-1} \epsilon_{e,-1} E_{57.6} t_{d,4}^{-1} f(p) d_{L,20}^{-2} \;
{\rm erg}\;{\rm cm}^{-2}\;{\rm s}^{-1}, \\
\varepsilon_m^{\rm SC} F_{\varepsilon_m}^{\rm SC} &\simeq& Y \varepsilon_m F_{\varepsilon_m},
\label{eq:sscflux_1}
\end{eqnarray}
where the luminosity distance $d_L$ is normalized by the value for $1+z=20$, 
$6.7\times10^{29}\;$cm.
In the case of significant KN effects, i.e., $\varepsilon_m^{\rm SC} > \Gamma_d \gamma_m m_e c^2/(1+z)$, 
the SSC emission is not important and $Y$ is smaller than the above value 
\citep[see][for details]{nakar09}.

The maximum energy of the electrons is determined by equating the acceleration timescale and 
the radiative cooling timescale. If the electron acceleration occurs in the whole region, 
we may estimate the (comoving) 
acceleration timescale as $t'_a = g 2\pi \gamma_e m_e c/(eB)$ where $g$ is a numerical factor.
For electrons with maximum Lorentz factor $\gamma_M$, the KN effect is typically significant, 
and $Y(\gamma_M) < 1$.  
Then we have $\gamma_M \simeq [3e/(g \sigma_T B)]^{1/2}$, and its characteristic synchrotron energy is
\begin{eqnarray}
\varepsilon_M &\simeq& \frac{3heB}{4\pi m_e c}\gamma_M^2 \frac{\Gamma_d}{1+z} 
= \frac{9h e^2}{4\pi m_e c \sigma_T g} \frac{\Gamma_d}{1+z} \nonumber \\
&\simeq& 1.8\times10^2\; g^{-1} E_{57.6}^{1/8} t_{d,4}^{-3/8} n_0^{-1/8} [(1+z)/20]^{-1} \; {\rm MeV}.
\label{eq:epsilon_M}
\end{eqnarray}

The energy $\varepsilon_a$ below which the synchrotron self-absorption effect is significant 
can be estimated by equating the synchrotron flux to the blackbody flux of the 
characteristic electrons in the shocked region, $F_{\varepsilon_a}
= [(1+z)^3/d_L^2] 2\pi m_e \gamma_{ch} (\varepsilon_a/h)^2 (r_d^2/\Gamma_d)$ 
\citep[e.g.,][]{mesAG97}.
The Lorentz factor $\gamma_{ch}$ of the characteristic electrons is given by $\gamma_a$ 
whose synchrotron energy is $\varepsilon_a$ in the case of $\gamma_a > \gamma_c$,
and otherwise by $\gamma_c$. Although more detailed calculations can be done by using 
the self-absorption coefficients \citep[e.g.,][]{melrose80,matsumiya03,toma08}, the 
above approximate derivation is sufficient for our current study. Thus we obtain
\begin{equation}
\gamma_a \simeq \cases{
69\; (1+Y)^{-1/6} \epsilon_{B,-2}^{-1/4} E_{57.6}^{-1/24} t_{d,4}^{1/8} n_0^{-1/24}, 
~~ {\rm for} ~~ \gamma_c < \gamma_a < \gamma_m, \cr
81\; (1+Y)^{1/2} \epsilon_{B,-2}^{7/20} E_{57.6}^{9/40} t_{d,4}^{1/8} n_0^{17/40},
~~ {\rm for} ~~ \gamma_a < \gamma_c < \gamma_m,
}
\end{equation}
The characteristic synchrotron energy of electrons with this Lorentz factor is given by
\begin{equation}
\varepsilon_a \simeq \cases{
1.5\times 10^{-3}\;(1+Y)^{-1/3} E_{57.6}^{1/6} t_{d,4}^{-1/2} n_0^{1/6} [(1+z)/20]^{-1}\;{\rm eV}, 
~~ {\rm for} ~~ \gamma_c < \gamma_a < \gamma_m, \cr
2.1 \times 10^{-3}\; (1+Y) \epsilon_{B,-2}^{6/5} E_{57.6}^{7/10} t_{d,4}^{-1/2} n_0^{11/10} [(1+z)/20]^{-1}\;
{\rm eV} ~~ {\rm for} ~~ \gamma_a < \gamma_c < \gamma_m.
}
\end{equation}

The high-energy absorption turnover energy $\varepsilon_{\gamma\gamma}$ due to the $e^+e^-$ pair 
creation can be estimated as follows \citep[e.g.,][]{zhang01,lithwick01}.  The photons 
are assumed to be roughly uniform over the emitting region and isotropic in its comoving 
frame, so that $\varepsilon_{at} \simeq \Gamma_d^2 m_e^2 c^4 / [\varepsilon_{\gamma\gamma} 
(1+z)^2]$ is the main target photon energy for the photons with energy 
$\varepsilon_{\gamma\gamma}$. 
Taking the spectral luminosity around $\varepsilon = \varepsilon_{at}$ as $L_\varepsilon = 
L_{\varepsilon_{at}} (\varepsilon/\varepsilon_{at})^{-\lambda}$, the total number of 
photons with $\varepsilon > \varepsilon_{at}$ is $N_{>\varepsilon_{at}} \simeq t_d 
\int_{\varepsilon_{at}}^{\infty} (L_\varepsilon/\varepsilon)d\varepsilon \simeq 
L_{\varepsilon_{at}} t_d$. 
By using $L_{\varepsilon_{at}} = 4\pi d_L^2 F_{\varepsilon_{at}}/(1+z)$, we obtain the 
opacity as $\tau_{\gamma\gamma}(\varepsilon_{\gamma\gamma}) \simeq (\sigma_T/10)
N_{>\varepsilon_{at}}/(4\pi r_d^2) = 1$, leading to
\begin{equation}
\varepsilon_{at} \simeq 9.0 \times 10^2\; (170)^{\frac{2.3-p}{p}} (1+Y)^{-\frac{2}{p}}
\epsilon_{e,-1}^{2-\frac{2}{p}} \epsilon_{B,-2}^{\frac{1}{2}-\frac{1}{p}}
E_{57.6}^{1/2} t_{d,4}^{-\frac{3}{2}+\frac{2}{p}} n_0^{\frac{1}{p}} {f(p)}^{2-\frac{2}{p}}
[(1+z)/20]^{-1}\; {\rm eV},
\end{equation}
where we have taken $\lambda = p/2$ since $\varepsilon_{at}$ is typically above 
$\varepsilon_m$.  The $\gamma\gamma$ self-absorption energy is then given by
\begin{equation}
\varepsilon_{\gamma\gamma} \simeq 6.9\; (170)^{\frac{p-2.3}{p}} (1+Y)^{\frac{2}{p}}
\epsilon_{e,-1}^{-2+\frac{2}{p}} \epsilon_{B,-2}^{-\frac{1}{2}+\frac{1}{p}}
E_{57.6}^{-\frac{1}{4}} t_{d,4}^{\frac{3}{4}-\frac{2}{p}} n_0^{-\frac{1}{4}-\frac{1}{p}}
{f(p)}^{-2+\frac{2}{p}} [(1+z)/20]^{-1}\; {\rm GeV}.
\label{eq:gammagamma}
\end{equation}
The condition for a significant absorption, $\varepsilon_{\gamma\gamma} < 
\varepsilon_m^{\rm SC}$, is rewritten as
\begin{equation}
\epsilon_{B,-2}^{1-\frac{1}{p}} n_0^{\frac{1}{p}} (1+Y)^{-\frac{2}{p}} > 
38\;(170)^{\frac{p-2.3}{p}} \epsilon_{e,-1}^{-6+\frac{2}{p}} E_{57.6}^{-1} 
t_{d,4}^{3-\frac{2}{p}} {f(p)}^{-6+\frac{2}{p}}.
\label{eq:condition_p}
\end{equation}
If this condition is satisfied and $\epsilon_e \gtrsim \epsilon_B$, the emission from the 
created pairs will significantly affect the observed spectrum.

Above we have assumed that the photon field is roughly uniform in the emitting region.  This may not be
valid, however, when electrons are in the fast-cooling regime (as is the case for our typical parameters).
If the acceleration process of electrons works only near the shock front, the emission at 
$\varepsilon \gtrsim \varepsilon_m$ is only produced in a thin layer from the shock front 
with a width of $\sim (\gamma_c/\gamma_m) r_d /\Gamma_d$ in the comoving frame. The 
annihilation process of the high-energy photons then mainly occurs outside the 
shocked region. In such a case the angles of the directions of motion of two given 
annihilating photons are typically very small, which significantly reduces the pair 
creation optical depth \citep{granot08}.  The $\gamma\gamma$ self-absorption energy 
$\varepsilon_{\gamma\gamma}$ could then be much larger than the above estimate by 
Equation~(\ref{eq:gammagamma}).  In addition to this,
the maximum synchrotron energy $\varepsilon_M$ is determined by the magnetic field
in the upstream region, instead of the emitting region \citep{li06}, which may be
smaller than the estimate by Equation~(\ref{eq:epsilon_M}) by a factor of $B_u \Gamma_d / B$,
where $B_u$ is the upstream field strength measured in its own frame.

An issue which has been a long-standing concern with the model is the possibility that the 
magnetic field amplification and the electron acceleration microphysics may be confined
to a region near the shock front.  Such small-scale magnetic field will decay within a 
couple of ion skin depths \citep[e.g.,][]{gruzinov01,kato05}, which may not explain the 
observed bright afterglows. On the other hand, the large-scale magnetic field amplified by 
macroscopic turbulence created by shock could survive over the whole emitting region, and
the observed afterglows may  be attributed to such large-scale field \citep{sironi07}. 
In this case the main acceleration process of electrons, e.g., second-order Fermi acceleration, 
could work uniformly in the shocked region, where the photons can be assumed to be roughly 
uniform and isotropic in the emitting region.  If this is the case, $\varepsilon_{\gamma\gamma}$
and $\varepsilon_M$ are given by the uniform photon field assumption, Equations~(\ref{eq:gammagamma})
and (\ref{eq:epsilon_M}), respectively.  We take this uniform photon field assumption in the 
rest of Appendix and in the main text.

\subsection{Case of Negligible Pair Production}
\label{subsec:case_np}

The parameter set of $E_{57.6} = t_{d,4} = n_0 = \epsilon_{B,-2} = \epsilon_{e,-1} = f(p) = 1$
satisfies the negligible pair production condition $\varepsilon_m^{\rm SC} < \varepsilon_{\gamma\gamma}$.
The radius and Lorentz factor 
of the shocked fluid are given by $r_d \simeq 2.8 \times 10^{18}\;$cm and $\Gamma_d 
\simeq 97$, respectively. We have $Y \simeq 2.7$, and then the characteristic electron 
Lorentz factors are $\gamma_m \simeq 4.1 \times 10^3$, $\gamma_a \simeq 55$, and
$\gamma_c \simeq 15$. The characteristic photon energies are
$\varepsilon_m \simeq 5.4\;$eV, $\varepsilon_a \simeq 9.7 \times 10^{-4}\;$eV,
$\varepsilon_m^{\rm SC} \simeq 1.8\times10^2\;$MeV ($< \Gamma_d \gamma_m m_e c^2 /(1+z) 
\simeq 10\;{\rm GeV}$), $\varepsilon_{\rm KN} \simeq 1.2\;$TeV,
$\varepsilon_{at} \simeq 2.9 \times 10^2\;$eV, and $\varepsilon_{\gamma\gamma} 
\simeq 21\;$GeV.  The flux normalization is given by $\varepsilon_m F_{\varepsilon_m} 
\simeq 4.4 \times 10^{-10}\; {\rm erg}\;{\rm cm}^{-2}\;{\rm s}^{-1}$, and 
$\varepsilon_m^{\rm SC} F_{\varepsilon_m}^{\rm SC}
\simeq 1.2\times 10^{-9}\;{\rm erg}\;{\rm cm}^{-2}\;{\rm s}^{-1}$.

The overall spectrum for this case is shown in Figure~\ref{fig:example_np}.
The synchrotron spectrum is approximately
\begin{equation}
\varepsilon F_{\varepsilon} \simeq \varepsilon_m F_{\varepsilon_m} \times \cases{
\left(\frac{\varepsilon_a}{\varepsilon_m}\right)^{1/2}
\left(\frac{\varepsilon}{\varepsilon_a}\right)^3, ~~ {\rm for}~~ \varepsilon < \varepsilon_a, \cr
\left(\frac{\varepsilon}{\varepsilon_m}\right)^{1/2}, ~~ {\rm for}~~ 
\varepsilon_a < \varepsilon < \varepsilon_m, \cr
\left(\frac{\varepsilon}{\varepsilon_m}\right)^{-\frac{p}{2}+1}, ~~ {\rm for}~~ 
\varepsilon_m < \varepsilon < \varepsilon_M.
}
\end{equation}
The SSC spectrum is approximately
\begin{equation}
\varepsilon F_{\varepsilon} \simeq \varepsilon_m^{\rm SC} F_{\varepsilon_m}^{\rm SC} \times \cases{
\left(\frac{\varepsilon_a}{\varepsilon_m^{\rm SC}}\right)^{1/2}
\left(\frac{\varepsilon}{\varepsilon_a}\right)^3, ~~ {\rm for}~~ \varepsilon < \varepsilon_a, \cr
\left(\frac{\varepsilon}{\varepsilon_m^{\rm SC}}\right)^{1/2}, ~~ {\rm for}~~ 
\varepsilon_a < \varepsilon < \varepsilon_m^{\rm SC}, \cr
\left(\frac{\varepsilon}{\varepsilon_m^{\rm SC}}\right)^{-\frac{p}{2}+1}, ~~ {\rm for}~~ 
\varepsilon_m < \varepsilon < \varepsilon_{\rm KN}, \cr
\left(\frac{\varepsilon_{\rm KN}}{\varepsilon_m^{\rm SC}}\right)^{-\frac{p}{2}+1} 
\left(\frac{\varepsilon}{\varepsilon_{\rm KN}}\right)^{-p+\frac{3}{2}}, ~~ {\rm for} ~~ 
\varepsilon > \varepsilon_{\rm KN},
}
\end{equation}
where we have neglected the segment of $\varepsilon F_{\varepsilon} \propto 
\varepsilon^{-p+2}$ at $\varepsilon > \varepsilon_{\rm KN}$ just for simplicity 
\citep[see][for more details]{nakar09}.  The SSC emission 
is absorbed by the pair creation within the emitting region 
at $\varepsilon > \varepsilon_{\gamma\gamma}$.  The spectral shape is given by the intrinsic
one multiplied by $1/\tau_{\gamma\gamma}(\varepsilon) \propto \varepsilon^{-p/2}$.

\subsection{Case of Significant Pair Production}
\label{subsec:case_sp}

The parameter set of $E_{57.6} = t_{d,4} = \epsilon_{B,-2} = f(p) = 1$, $n_0 = 10^2$, and 
$\epsilon_{e,-1} = 2$ satisfies the significant pair production condition 
$\varepsilon_{\gamma\gamma} < \varepsilon_m^{\rm SC}$.
The radius and Lorentz factor of the shocked fluid at $t_d$ are $r_d \simeq 8.9 \times 10^{17}
\;$cm and $\Gamma_d \simeq 55$, respectively. The characteristic Lorentz factors of the 
electrons are $\gamma_m \simeq 4.6 \times 10^3$, $\gamma_a \simeq 42$, and $\gamma_c 
\simeq 0.55$, where we have used the value of $Y$ calculated below.  The cooling Lorentz 
factor $\gamma_c < 1$ would imply that the electron energy distribution has a bump at 
$\gamma_e \simeq 1$. The synchrotron self-absorption effect may be so strong that
the electrons at $\gamma_e \simeq 1$ can be reheated to higher energies. In any case, 
however, the electron energy distribution above $\gamma_a$ is not affected, and the 
synchrotron spectrum below $\varepsilon_a$ is not relevant.
The characteristic photon energies are $\varepsilon_m \simeq 21\;$eV,
$\varepsilon_a \simeq 1.8 \times 10^{-3}\;$eV, $\varepsilon_m^{\rm SC} \simeq 9.2\times10^2\;$MeV
($< \Gamma_d \gamma_m m_e c^2/(1+z) \simeq 6.5\;$GeV),
$\varepsilon_{\rm KN} \simeq 91\;$GeV, $\varepsilon_{at} \simeq 3.2 \times 10^3\;$eV,
and $\varepsilon_{\gamma\gamma} \simeq 6.1\times10^2\;$MeV. The flux normalizations are given by
$\varepsilon_m F_{\varepsilon_m} \simeq 5.8 \times 10^{-10}\;{\rm erg}\;{\rm cm}^{-2}\;
{\rm s}^{-1}$ and $\varepsilon_m^{\rm SC} F_{\varepsilon_m}^{\rm SC} \simeq 2.7 \times 
10^{-9}\;{\rm erg} \;{\rm cm}^{-2}\;{\rm s}^{-1}$.

The pair creation opacity is larger than unity for photons with $\varepsilon_m^{\rm SC}$,
so most of the SSC radiation luminosity will be converted into $e^+e^-$ pairs.
If the compactness parameter $l'$ in the shocked region is as high as $\sim 10^2$, 
the radiation will be thermalized by Compton scattering on the pairs, resulting in a Wien 
spectrum \citep{peer04}.  In our case,
\begin{equation}
l' = \frac{\sigma_T \epsilon_e E_{\rm iso} t_d^{-1}}{8\pi m_e c^3 \Gamma_d^3 r_d}
\simeq 0.6\; \left(\frac{\epsilon_e}{0.2}\right) E_{57.6}^{3/8} t_{d,4}^{-1/8} 
\left(\frac{n}{10^2}\right)^{5/8},
\label{eq:compactness}
\end{equation}
which is high compared to typical cases of Pop.~I/II afterglows, but still not as 
high as $10^2$.  We can approximately derive non-thermal synchrotron and inverse 
Compton emission spectra by considering the following cascade process.

The Lorentz factors of the created pairs are determined by $\gamma_\pm m_e c^2 \approx
(1+z) \varepsilon \Gamma_d^{-1}/2$. The indices of the injected energy distribution of 
the pairs are the same as those of the photons with $\varepsilon \geq 
\varepsilon_{\gamma\gamma}$, being $dn_\pm/d\gamma_\pm \propto \gamma_\pm^{-3/2}$ for 
$\gamma_{\pm,m} < \gamma_\pm < \gamma_{\pm,p}$,
$\propto \gamma_\pm^{-(p/2)-1}$ for $\gamma_{\pm,p} < \gamma_\pm < \gamma_{\pm,k}$, and 
$\propto \gamma_\pm^{-p-1/2}$ for $\gamma_\pm > \gamma_{\pm,k}$, where
\begin{equation}
\gamma_{\pm,m} \simeq \frac{\varepsilon_{\gamma\gamma}(1+z)}{2 \Gamma_d m_e c^2}, ~~~
\gamma_{\pm,p} \simeq \frac{\varepsilon_m^{\rm SC}(1+z)}{2 \Gamma_d m_e c^2}, ~~~
\gamma_{\pm,k} \simeq \frac{\varepsilon_{\rm KN}(1+z)}{2 \Gamma_d m_e c^2}. 
\end{equation}
For the parameter values adopted here we have $\gamma_{\pm,m} \simeq 220$, $\gamma_{\pm,p} 
\simeq 330$, and $\gamma_{\pm,k} \simeq 3.2 \times 10^4$. The corresponding synchrotron 
energies are $\varepsilon_{\pm,m} \simeq 4.7 \times 10^{-2}\;$eV, $\varepsilon_{\pm,p} 
\simeq 0.11\;$eV, and $\varepsilon_{\pm,k} \simeq 1.0\;$keV, respectively. 
Since the synchrotron and SSC cooling is very fast, the averaged energy distribution 
of the pairs over the dynamical timescale is given by
\begin{equation}
\frac{dn_\pm}{d\gamma_\pm} \propto \cases{
\gamma_\pm^{-2}, ~~ {\rm for} ~~ \gamma_{\pm,c} < \gamma_\pm < \gamma_{\pm,m}, \cr
\gamma_\pm^{-5/2}, ~~ {\rm for} ~~ \gamma_{\pm,m} < \gamma_\pm < \gamma_{\pm,p}, \cr
\gamma_\pm^{-(p/2)-2}, ~~ {\rm for} ~~ \gamma_{\pm,p} < \gamma_\pm < \gamma_{\pm,k}, \cr 
\gamma_\pm^{-p-(3/2)}, ~~ {\rm for} ~~ \gamma_\pm > \gamma_{\pm,k}.
}
\end{equation}
The shapes of the synchrotron and inverse Compton emission spectra of the pairs are 
obtained straightforwardly. The SSC emission of the pairs peaks at 
$\varepsilon_{\pm,p}^{\rm SC} \simeq 2 \gamma_{\pm,p}^2 \varepsilon_{\pm,p} \simeq 23
\;$keV, and the pairs IC-scatter the original synchrotron emission, which peaks at 
$\varepsilon_{\pm,p}^{\rm IC} \simeq 2 \gamma_{\pm,p}^2 \varepsilon_m \simeq 4.5\;$MeV.
The KN effect is not significant since $\varepsilon_{\pm,p}^{\rm SC}, 
\varepsilon_{\pm,p}^{\rm IC} < \Gamma_d \gamma_{\pm,p} m_e c^2/(1+z) \simeq 4.6\times10^2\;$MeV.
The original electrons IC-scatter the pair synchrotron emission, which 
peaks at $\varepsilon_m^{\rm IC} \simeq 2 \gamma_m^2 \varepsilon_{\pm,p} = 
\varepsilon_{\pm,p}^{\rm IC}$.

In order to determine the flux normalization of the various emission components, we define
\begin{eqnarray}
&&\frac{\varepsilon_m^{\rm SC} F_{\varepsilon_m}^{\rm SC}}{\varepsilon_m F_{\varepsilon_m}} = Y^{\rm SC}, ~~
\frac{\varepsilon_m^{\rm IC} F_{\varepsilon_m}^{\rm IC}}{\varepsilon_m F_{\varepsilon_m}} = Y^{\rm IC}, ~~ 
\frac{\varepsilon_{\pm,p} F_{\varepsilon_{\pm,p}}}{\varepsilon_m F_{\varepsilon_m}} = x, \\ 
&&\frac{\varepsilon_{\pm,p}^{\rm SC} F_{\varepsilon_{\pm,p}}^{\rm SC}}
{\varepsilon_{\pm,p} F_{\varepsilon_{\pm,p}}} = Y^{\rm SC}_{\pm}, ~~
\frac{\varepsilon_{\pm,p}^{\rm IC} F_{\varepsilon_{\pm,p}}^{\rm IC}}
{\varepsilon_{\pm,p} F_{\varepsilon_{\pm,p}}} = Y^{\rm IC}_{\pm}.
\end{eqnarray}
Here we cannot use Equations~(\ref{eq:y_1}) and (\ref{eq:sscflux_1}), using instead the 
following relations for the scattering by the original electrons,
\begin{eqnarray}
Y^{\rm SC} = \frac{Y^{\rm IC}}{x} \simeq \tau \gamma_c \gamma_m \frac{p-1}{p-2} \simeq
\frac{\epsilon_e}{\epsilon_B} \frac{1}{1+Y},
\label{eq:ysc}
\end{eqnarray}
where $Y = Y^{\rm SC} + Y^{\rm IC}$. 

The total number of the pairs can be estimated as
\begin{eqnarray}
N_\pm &\sim& 2t_d \int_{\varepsilon_{\gamma\gamma}}^{\infty} 
\frac{L_\varepsilon^{\rm SC}}{\varepsilon} d\varepsilon
\simeq 2 L_{\varepsilon_{\gamma\gamma}}^{\rm SC} t_d = 8\pi d_L^2 F_{\varepsilon_m}^{\rm SC}
\left(\frac{\varepsilon_{\gamma\gamma}}{\varepsilon_m^{\rm SC}}\right)^{-1/2} \frac{t_d}{1+z} \nonumber \\
&\simeq& \frac{2 \epsilon_e E_{\rm iso}}{\varepsilon_m^{\rm SC} (1+z)} 
\left(\frac{\varepsilon_{\gamma\gamma}}{\varepsilon_m^{\rm SC}}\right)^{-1/2} \frac{p-2}{p-1} 
\frac{Y^{\rm SC}}{1+Y}.
\end{eqnarray}
On the other hand, the number of the original electrons is
$N \sim 4\pi r_d^3 n = E_{\rm iso}/(\Gamma_d^2 m_p c^2)$. These two equations lead to
\begin{equation}
\frac{N_\pm}{N} \sim \frac{2 \Gamma_d \gamma_m m_e c^2}{\varepsilon_m^{\rm SC} (1+z)}
\left(\frac{\varepsilon_{\gamma\gamma}}{\varepsilon_m^{\rm SC}}\right)^{-1/2} \frac{Y^{\rm SC}}{1+Y}.
\label{eq:n_ratio}
\end{equation}
The flux ratio $Y_{\pm}^{\rm SC}$ satisfies the following relation
\begin{eqnarray}
Y_\pm^{\rm SC} &=& Y^{\rm IC}_{\pm} x
\simeq \sigma_T \frac{r_d}{\Gamma_d} \int_{\gamma_{\pm,c}}^{\infty} d\gamma_\pm 
\frac{dn_\pm}{d\gamma_\pm} \gamma_\pm^2 \nonumber \\
&\simeq& 2 \tau \frac{N_\pm}{N} \gamma_{\pm,c} \gamma_{\pm,m}^{1/2} \gamma_{\pm,p}^{1/2} 
\frac{p-1}{p-2} 
\simeq 2 \tau \gamma_{\pm,c} \gamma_m \frac{p-1}{p-2} \frac{Y^{\rm SC}}{1+Y} \nonumber \\
&=& 2 Y^{\rm SC} \frac{\gamma_{\pm,c}}{\gamma_c} \frac{Y^{\rm SC}}{1+Y}
= \frac{2 (Y^{\rm SC})^2}{1+Y_\pm},
\label{eq:ypmsc}
\end{eqnarray}
where $Y_\pm = Y_\pm^{\rm SC} + Y_\pm^{\rm IC}$, and we have used 
$\gamma_{\pm,c}/\gamma_c = (1+Y)/(1+Y_\pm)$.  The number ratio $N_\pm/N$ is equal to 
the ratio of the synchrotron fluxes of the pairs and the original electrons, 
$F_{\varepsilon_{\pm,c}}/F_{\varepsilon_c}$. Thus, the ratio of the 
$\varepsilon F_{\varepsilon}$ fluxes is
\begin{eqnarray}
x &=& \frac{\gamma_{\pm,p}^2 F_{\varepsilon_{\pm,c}} (\varepsilon_{\pm,m}/\varepsilon_{\pm,c})^{-1/2}
(\varepsilon_{\pm,p}/\varepsilon_{\pm,m})^{-3/4}}{\gamma_m^2 F_{\varepsilon_c} 
(\varepsilon_m/\varepsilon_c)^{-1/2}} \nonumber \\
&=& \frac{Y^{\rm SC}}{1+Y_\pm}.
\label{eq:x}
\end{eqnarray}
Equations~(\ref{eq:ysc}), (\ref{eq:ypmsc}), and (\ref{eq:x}) reduce to 
\begin{equation}
\frac{x(1-x-x^2)}{(1-2x-2x^2)^2} = \frac{\epsilon_e}{\epsilon_B}, ~~ Y^{\rm SC} = \frac{x}{1-2x-2x^2},
~~ Y^{\rm SC}_\pm = 2x Y^{\rm SC},
\end{equation}
which lead to a solution $x \simeq 0.34$, $Y^{\rm SC} \simeq 3.5$, $Y^{\rm IC} \simeq 1.2$, 
$Y \simeq 4.7$, $Y^{\rm SC}_\pm \simeq 2.4$, $Y^{\rm IC}_\pm \simeq 7.1$, and 
$Y_\pm \simeq 9.5$.
These factors are only dependent on $\epsilon_B$ and $\epsilon_e$, being constant as long as 
$\gamma_{\pm,c} < \gamma_{\pm,m} < \gamma_{\pm,p} < \gamma_{\pm,k}$ is satisfied.

Figure~\ref{fig:example_sp} shows an approximate spectrum of the emission by the original 
electrons and the pairs. The synchrotron spectrum of the pairs is approximated as
\begin{equation}
\varepsilon F_{\varepsilon} \simeq \varepsilon_{\pm,p} F_{\varepsilon_{\pm,p}} \times \cases{
\left(\frac{\varepsilon_{\pm,m}}{\varepsilon_{\pm,p}}\right)^{1/4}
\left(\frac{\varepsilon_{a}}{\varepsilon_{\pm,m}}\right)^{1/2}
\left(\frac{\varepsilon}{\varepsilon_{a}}\right)^3, ~~ {\rm for}~~ \varepsilon < \varepsilon_{a}, \cr
\left(\frac{\varepsilon_{\pm,m}}{\varepsilon_{\pm,p}}\right)^{1/4}
\left(\frac{\varepsilon}{\varepsilon_{\pm,m}}\right)^{1/2}, ~~ {\rm for} ~~
\varepsilon_{a} < \varepsilon < \varepsilon_{\pm,m}, \cr
\left(\frac{\varepsilon}{\varepsilon_{\pm,p}}\right)^{1/4}, ~~ {\rm for}~~ 
\varepsilon_{\pm,m} < \varepsilon < \varepsilon_{\pm,p}, \cr
\left(\frac{\varepsilon}{\varepsilon_{\pm,p}}\right)^{-\frac{p}{4}+\frac{1}{2}}, ~~ {\rm for}~~ 
\varepsilon_{\pm,p} < \varepsilon < \varepsilon_{\pm,k}, \cr
\left(\frac{\varepsilon_{\pm,k}}{\varepsilon_{\pm,p}}\right)^{-\frac{p}{4}+\frac{1}{2}}
\left(\frac{\varepsilon}{\varepsilon_{\pm,k}}\right)^{-\frac{p}{2}+\frac{3}{4}}, ~~ {\rm for} ~~
\varepsilon > \varepsilon_{\pm,k}.
}
\end{equation}
The SSC spectrum of the pairs is
\begin{equation}
\varepsilon F_{\varepsilon} \simeq \varepsilon_{\pm,p}^{\rm SC} F_{\varepsilon_{\pm,p}}^{\rm SC} \times \cases{
\left(\frac{\varepsilon_{\pm,m}^{\rm SC}}{\varepsilon_{\pm,p}^{\rm SC}}\right)^{1/4}
\left(\frac{\varepsilon_{a}}{\varepsilon_{\pm,m}^{\rm SC}}\right)^{1/2}
\left(\frac{\varepsilon}{\varepsilon_{a}}\right)^3, ~~ {\rm for}~~ \varepsilon < \varepsilon_{a}, \cr
\left(\frac{\varepsilon_{\pm,m}^{\rm SC}}{\varepsilon_{\pm,p}^{\rm SC}}\right)^{1/4}
\left(\frac{\varepsilon}{\varepsilon_{\pm,m}^{\rm SC}}\right)^{1/2}, ~~ {\rm for} ~~
\varepsilon_{a} < \varepsilon < \varepsilon_{\pm,m}^{\rm SC}, \cr
\left(\frac{\varepsilon}{\varepsilon_{\pm,p}^{\rm SC}}\right)^{1/4}, ~~ {\rm for}~~ 
\varepsilon_{\pm,m}^{\rm SC} < \varepsilon < \varepsilon_{\pm,p}^{\rm SC}, \cr
\left(\frac{\varepsilon}{\varepsilon_{\pm,p}^{\rm SC}}\right)^{-\frac{p}{4}+\frac{1}{2}}, ~~ {\rm for}~~ 
\varepsilon_{\pm,p}^{\rm SC} < \varepsilon < \varepsilon_{\pm,k}^{\rm SC}, \cr
\left(\frac{\varepsilon_{\pm,k}^{\rm SC}}{\varepsilon_{\pm,p}^{\rm SC}}\right)^{-\frac{p}{4}+\frac{1}{2}}
\left(\frac{\varepsilon}{\varepsilon_{\pm,k}^{\rm SC}}\right)^{-\frac{p}{2}+\frac{3}{4}}, ~~ {\rm for} ~~
\varepsilon > \varepsilon_{\pm,k}^{\rm SC},
}
\end{equation}
where $\varepsilon_{\pm,m}^{\rm SC} \simeq 2 \gamma_{\pm,m}^2 \varepsilon_{\pm,m}$ and 
$\varepsilon_{\pm,k}^{\rm SC} \simeq 2 \gamma_{\pm,k}^2 \varepsilon_{\pm,k}$.
The IC spectrum of the pairs is
\begin{equation}
\varepsilon F_{\varepsilon} \simeq \varepsilon_{\pm,p}^{\rm IC} F_{\varepsilon_{\pm,p}}^{\rm IC} \times \cases{
\left(\frac{\varepsilon_{\pm,m}^{\rm IC}}{\varepsilon_{\pm,p}^{\rm IC}}\right)^{1/4}
\left(\frac{\varepsilon_{a}}{\varepsilon_{\pm,m}^{\rm IC}}\right)^{1/2}
\left(\frac{\varepsilon}{\varepsilon_{a}}\right)^3, ~~ {\rm for}~~ \varepsilon < \varepsilon_{a}, \cr
\left(\frac{\varepsilon_{\pm,m}^{\rm IC}}{\varepsilon_{\pm,p}^{\rm IC}}\right)^{1/4}
\left(\frac{\varepsilon}{\varepsilon_{\pm,m}^{\rm IC}}\right)^{1/2}, ~~ {\rm for} ~~
\varepsilon_{a} < \varepsilon < \varepsilon_{\pm,m}^{\rm IC}, \cr
\left(\frac{\varepsilon}{\varepsilon_{\pm,p}^{\rm IC}}\right)^{1/4}, ~~ {\rm for}~~ 
\varepsilon_{\pm,m}^{\rm IC} < \varepsilon < \varepsilon_{\pm,p}^{\rm IC}, \cr
\left(\frac{\varepsilon}{\varepsilon_{\pm,p}^{\rm IC}}\right)^{-\frac{p}{4}+\frac{1}{2}}, ~~ {\rm for}~~ 
\varepsilon_{\pm,p}^{\rm IC} < \varepsilon < \varepsilon_{\pm,k}^{\rm IC}, \cr
\left(\frac{\varepsilon_{\pm,k}^{\rm IC}}{\varepsilon_{\pm,p}^{\rm IC}}\right)^{-\frac{p}{4}+\frac{1}{2}}
\left(\frac{\varepsilon}{\varepsilon_{\pm,k}^{\rm IC}}\right)^{-\frac{p}{2}+1}, ~~ {\rm for} ~~
\varepsilon > \varepsilon_{\pm,k}^{\rm IC},
}
\end{equation}
where $\varepsilon_{\pm,m}^{\rm IC} \simeq 2 \gamma_{\pm,m}^2 \varepsilon_m$ and 
$\varepsilon_{\pm,k}^{\rm IC} \simeq 2 \gamma_{\pm,k}^2 \varepsilon_m$.
The IC spectrum of the original electrons is the same of that of the pairs except for 
the flux normalization.  We have superposed these two IC components in 
Figure~\ref{fig:example_sp}.
The energies $\varepsilon_a$ and $\varepsilon_{\gamma\gamma}$ should be 
re-calculated by taking into account the emission of the pairs. For the adopted parameters, 
however, we found that $\varepsilon_a$ is larger, and $\varepsilon_{\gamma\gamma}$ is smaller,
than the original values only by a factor of $\simeq 2$, which do not significantly affect 
the overall spectrum.
We have $\varepsilon_{\gamma\gamma} > \varepsilon_{\pm,p}^{\rm IC}$, so that no further 
significant cascade emission is expected for the parameters adopted here.

We have assumed that 
the energies of the pairs created in the emitting region are just given by the annihilated 
photon energies and the shock acceleration of the energetic pairs is not effective.
The total pair energy density is comparable to the total injected energy density 
of the original electrons, while the ratio of the number densities is given by Equation 
(\ref{eq:n_ratio}).  Thus the characteristic Lorentz factor achieved by the pairs from their shock
acceleration is estimated as $\gamma_{\pm,ac} \simeq \gamma_m N/N_\pm \sim \gamma_{\pm,p} 
(\varepsilon_{\gamma\gamma}/\varepsilon_m^{\rm SC})^{1/2} [(1+Y)/Y^{\rm SC}] \sim 
\gamma_{\pm,p}$.  This implies that the shock acceleration does not affect significantly the pair 
energy distribution given from the annihilating photon energy distribution.

\end{document}